\documentclass[aps,nofootinbib,superscriptaddress, showpacs,preprintnumbers,  nofootinbibt,twocolumn]{revtex4-2}

\usepackage{epsfig}
\usepackage{multirow}
\usepackage{eurosym}
\usepackage{dcolumn}
\usepackage{bm}
\usepackage{enumerate}
\usepackage{float}
\usepackage{epstopdf}
\usepackage{amsmath}
\usepackage{bm}
\usepackage{amsfonts}
\usepackage{amssymb}
\usepackage{graphicx}
\usepackage{alphalph,mathtools}
\usepackage{etoolbox}
\usepackage{color}
\usepackage{booktabs}
\usepackage{footnote}
\usepackage{makecell,tabularx}

\usepackage{hyperref}
\hypersetup{colorlinks,citecolor=blue}
\hypersetup{colorlinks=true,linkcolor=magenta,filecolor=magenta,    urlcolor=blue}

\def\be{\begin{equation}}
\def\ee{\end{equation}}
\def\bea{\begin{eqnarray}}
\def\eea{\end{eqnarray}}

\begin{document}
\color{black}       
%
\title{ \bf Cosmic evolution in $f(Q,T)$ gravity: Exploring a higher-order time-dependent function of deceleration parameter with observational constraints}

\author{Madhur khurana}
\email{K.madhur2000@gmail.com}
\affiliation{Department of Applied Physics, Delhi Technological University, Delhi-110042, India}
\author{Himanshu Chaudhary}
\email{himanshuch1729@gmail.com}
\affiliation{Department of Applied Mathematics, Delhi Technological University, Delhi-110042, India} 
\affiliation{Pacif Institute of Cosmology and Selfology (PICS), Sagara, Sambalpur 768224, Odisha, India}
\affiliation{Department of Mathematics, Shyamlal College, University of Delhi, Delhi-110032, India.}
\author{Saadia Mumtaz}
\email{saadia.icet@pu.edu.pk (Corresponding Author)}
\affiliation{Institute of Chemical Engineering and Technology, University of the Punjab, Quaid-e-Azam Campus, Lahore-54590, Pakistan}
\author{S. K. J. Pacif}
\email{shibesh.math@gmail.com}
\affiliation{Pacif Institute of Cosmology and Selfology (PICS), Sagara, Sambalpur 768224, Odisha, India}
\author{G. Mustafa}
\email{gmustafa3828@gmail.com (Corresponding Author)}
\affiliation{Department of Physics,
Zhejiang Normal University, Jinhua 321004, People’s Republic of China,}
\affiliation{New Uzbekistan University, Movarounnahr street 1, Tashkent 100000, Uzbekistan}

\begin{abstract}
In this research paper, we explore a well-motivated parametrization of the time-dependent deceleration parameter, characterized by a cubic form in cosmic time $t$, in the context of late-time cosmic acceleration. The whole analysis of our work is within the framework of $f(Q,T)$ gravity theory, where we consider the background metric as the homogeneous and isotropic one represented by the Friedmann–Lemaître–Robertson–Walker (FLRW) metric. We solve the field equations with the considered functional form of $d(t)$. In order to find some best fit values of the model parameters and validate our obtained model, we constrain the model using some recent observational datasets, including cosmic chronometer (CC), Supernovae (SNIa), Baryon Acoustic Oscillation (BAO), Cosmic Microwave Background Radiation (CMB), Gamma-ray Burst (GRB), and Quasar (Q) datasets. The joint analysis of these datasets results in tighter constraints for the model parameters, enabling us to discuss both the physical and geometrical aspects of the model. Moreover, we determine the present values of the deceleration parameter ($q_0$), the Hubble parameter ($H_0$), and the transition redshift ($z_t$) from deceleration to acceleration ensuring consistency with some recent results of Planck 2018. Our statistical analysis yields highly improved results, surpassing those obtained in previous investigations. Overall, this study presents valuable insights into the higher order $q(t)$ model and its implications for late-time cosmic acceleration, shedding light on the nature of the late universe.
\end{abstract}

\maketitle
\tableofcontents

\newpage
\section{\label{sec1}Introduction}
In contemporary times, one of the most significant breakthroughs is
the discovery of cosmic accelerated expansion, which has been
confirmed through a range of observational techniques \cite{1,2}.
This expansion is accompanied by a mysterious energy component known
as "dark energy" (DE), characterized by substantial negative
pressure. Despite its enigmatic nature, DE constitutes about $70\%$
of the cosmic content and plays a crucial role in maintaining the
overall energy density of the universe consistent with predictions
from inflationary theory. The investigation into the dominant
constituents of the universe, namely dark energy and dark matter
(DM), stands as a formidable challenge in modern physics,
representing nearly $95\%$ of the universe's imperceptible
composition. Dark matter, an unseen type of matter exhibiting weak
interactions with electromagnetic radiation, is detectable primarily
through its gravitational effects on nearby ordinary matter. The
existence of DM is supported by various observable phenomena,
including rotation curves and mass discrepancies.\\\\
In the absence of any substantial evidence supporting the dark
sources, alternative avenues have been explored extensively. The
investigation of enigmatic approaches to these exotic terms has been
delineated through two distinct methods: modifying matter sources or
introducing additional degrees of freedom into the gravitational
action. The initial approach involves the alteration of the matter
sector within the Einstein-Hilbert Lagrangian density through the
incorporation of various proposals. Numerous models have been
proposed to explain DE, some of which correlate with observational
evidence. Among these, the $\Lambda$CDM model is widely embraced due
to its alignment with observations, although it also presents
certain ambiguities such as fine-tuning and coincidence issues
\cite{fine-tuning1,fine-tuning2,coincidence1,coincidence2,coincidence3}. To address these limitations, alternative DE models
like quintessence \cite{Quintessence1,Quintessence2,Quintessence3}, phantom \cite{Phantom}, $k$-essence \cite{k-inflation1,quintessence4,k-essence}, and Chaplygin
gas \cite{Chaplygin} have been introduced, aiming to provide effective
explanations for various cosmic inquiries and to explore diverse
aspects of this exotic energy. Local (model-independent) $H_{0}$ values are biased to larger values than Planck $\Lambda$CDM. This disfavors Quintessence, K-essence, and any DE model in the Quintessence regime $w_{DE}(z) > -1$, while favoring phantom models, \cite{Colgain,Colgain2,d2023cosmographic,vagnozzi2023seven,vagnozzi2020new} $H_{0}$ tension is telling us new things about alternative DE models. Preceding the cosmic acceleration
phase, the universe underwent a deceleration phase during its early
epochs, where the impact of DE was relatively minor. It is believed
that density perturbations during this period played a pivotal role
in shaping cosmic structures. As a result, understanding the
complete evolutionary timeline requires a cosmological model capable
of describing both acceleration and deceleration phases.\\\\
The second approach involves an extension of the
gravitational component in GR by introducing a DE source while
keeping the matter sector unchanged. As the first category holds
intriguing implications, it has not gained as much support due to
certain ambiguities. In contrast, modified gravity frameworks have
proven to be quite valuable due to their effective execution in the
field of cosmology. Einstein formulated the concept of
geometry-matter coupling whose adaptation has been integral to the
development of General Relativity (GR). Several alternative theories
of gravity, which incorporate additional curvature terms in the
gravitational action, have been thoroughly examined in the
scientific literature. Some well-known examples of these modified
gravity theories encompass$f(R)$ gravity \cite{New1}, Gauss-Bonnet
gravity \cite{New2}, $f(R,T)$ gravity \cite{New3}, the scalar-tensor
theory \cite{New4}, $f(T)$ gravity \cite{New5}, $f(T, T_{G})$ gravity
\cite{mustafa2019wormhole}, $f(Q)$ gravity \cite{New7}, etc. These alterations are rooted
in the metric tensor $g_{ij}$ being considered a dynamic variable.
Interestingly, an alternative approach to GR has been gaining
momentum in the scientific literature, known as $f(Q,T)$ gravity
\cite{New8} appearing as an extension of symmetric teleparallel
gravity, where $Q$ and $T$ correspond to the non-metricity and the
trace of the energy-momentum tensor, respectively. This novel theory
has sparked significant interest in investigating the late universe
and has been the subject of different researchers in various
contexts \cite{New9,New10,New11,New12}.\\\\
The Einstein field equations within the framework of General Relativity (GR) are notoriously intricate, comprising a set of nonlinear differential equations that pose significant challenges in terms of finding analytical solutions. To simplify these complexities, physicists often make certain physical assumptions, such as establishing relationships between the pressure and energy density of the universe's contents. However, when introducing components like dark energy, these equations become even more tangled. This complexity also extends to modified gravity (MG) theories, which incorporate higher-order derivative terms. To obtain tractable solutions for the field equations, whether within GR or MG theories, researchers employ a variety of techniques, including dynamical system analysis, autonomous systems, and notably, the model-independent approach. The model-independent approach, which typically involves the parametrization of any cosmological parameter, which is generally a functional form of the parameter. While the parametrization of cosmological parameters may initially appear adhoc, it proves to be a legitimate and powerful method when considering the broad spectrum of possible evolutions for the geometrical and physical parameters from a mathematical standpoint. This idea has garnered increasing interest in the realm of mathematical cosmology.  This concept of cosmological parametrization has been extensively explored in the scientific literature, with comprehensive discussions by Pacif \cite{29,pacif2017reconstruction}, providing valuable insights into its application and significance in cosmological studies. There are also various studies on the model-independent approach by Eric V. Linder within some cosmological contexts. Specifically, he delved into the intricate studies of cosmological parametrization, a crucial facet of his research, which aimed to comprehensively study and understand a wide range of dark energy models that play a pivotal role in understanding the universe's accelerating expansion \cite{linder2003exploring,linder1997cosmological,linder2005many,linder2008dynamics,linder2008mapping}\\\\\
Modern cosmology advocates the investigation of kinematic
quantities, often referred to as "Cosmography" or "Cosmo-kinetics."
This approach relies on observational data and sidesteps prior
assumptions about gravity theory or specific cosmological models.
Cosmography's adoption of symmetry principles offers a means to
navigate debates surrounding DE, dark matter, and related topics
without invoking Einstein field equations (Friedmann equations).
While cosmography does not directly involve the scale factor, it
allows for some inference into its evolutionary history. The
foundation of modern cosmology traces back to the work by Sandage
\cite{sandage1998beginnings}, who introduced fundamental cosmographic parameters. These
parameters include the Hubble parameter ($H_0$), which dictates the
rate of cosmic expansion, and the deceleration parameter ($q_0$),
responsible for accounting for the effects of gravity that slow down
the expansion rate. 
Despite the emergence of DE altering the
cosmological landscape, the dynamics of cosmic evolution remain
intimately connected to the deceleration parameter ($q$), defined as
$q=-\frac{a\ddot{a}}{\dot{a}^2}$, where $a(t)$ represents the scale
factor. A positive value of $q$ ($\ddot{a}<0$) indicates
deceleration, while a negative value implies acceleration. The
Hubble parameter characterizes the linear temporal evolution of
$a(t)$, and a non-linear correction term ($q_0$) introduces local
instabilities and chaotic behavior \cite{bolotin2015cosmology}. The deceleration
parameter ($q$) also plays a vital role in examining the dynamics of
observable galaxy numbers. Dark energy gains empirical support
through various modifications to fit observational data, making its
parameterization as a function of the scale factor ($a(t)$) or
redshift ($z$) a valuable approach.\\\\
In the current paper, we have considered a parametrization of the deceleration parameter given in \cite{sofuouglu2023observational} within the classical gravity and extend the analysis in $f(Q,T)$ gravity, which is also a special case of the parametrization considered in \cite{29}. A plethora of parametric forms for the deceleration parameter have been explored, each with its own set of limitations. While some diverge at large cosmic times, others remain well-behaved at low
redshifts ($z<<1$) \cite{capozziello1,capozziello2,capozziello5,Capozziello3,Capozziello4,Orlando1,Orlando2,Orlando3,Orlando4,Orlando5,extra1,extra2,extra3,extra4,extra5,20:2012ya,article21,22:2008mt,23:2004lze,24:2007gvk,25:2009zza,articl26,27:2013lya,article28,29:2006gs,30:2001mx,31:2015ali,q(z)1,q(z)2,q(z)3,q(z)4,q(z)5,q(z)6,chaudhary2023cosmological,Modelindependent,mamon2017parametric,mamon2018constraints,khurana2023cosmological,khurana2023exploring}. The adoption of parametric assumptions can occasionally lead to misinterpretations of the nature of DE, prompting an interest in non-parametric models that directly infer evolutionary mechanisms from observational data
\cite{13,14,15,16,17}. Nevertheless, the parameterization of $q$ proves to be a more effective strategy for investigating the
transition from cosmic deceleration to acceleration.\\\\
Highlighting the versatility and gravitational-theory independence of the parameterized $q$ approach. In this study, we have made some analysis of the model produced by the cubic parametrization of the deceleration parameter and found tighter constraints to the model parameters involved. Additionally, we have done some cosmographic tests in order to differentiate a better-performing model when compared with cosmological data.\\\\\
The paper is organized in the following manner. In section \ref{sec1}, we provide an overview of the current state of modern cosmology, highlighting the various scenarios proposed to explain the late-time cosmic acceleration of the universe. We also outline the scope and objectives of the paper. In the section \ref{sec2}, we introduce the $f(Q,T)$ gravity framework and present the associated field equations. We derive the modified Friedmann equations within this context. In section \ref{sec3}, we discuss a model-independent approach to parametrize the cosmological model. This section lays the foundation for our subsequent analyses. Section \ref{sec4}, focuses on characterizing the model through dynamical variables, providing insights into its behavior and evolution. In Section \ref{sec5}, we derive all cosmological and physical parameters in terms of redshift, enhancing our understanding of the model's evolution. Section \ref{sec6} employs the Markov chain Monte Carlo (MCMC) method to constrain the model's parameters. We determine the best-fit values of model parameters by comparing them with the present Hubble rate function across different datasets. In Section \ref{sec7}, we validate the model's predictions by comparing them with observational data, Section \ref{sec8} delves into the kinematic cosmographic parameters, including the deceleration and jerk parameters, providing a detailed analysis of their implications within the model. Sections \ref{sec9} and \ref{sec10} are dedicated to the statefinder and $Om$ diagnostic tests, offering further insights into the model's behavior and compatibility with observational data. In \ref{sec11}, we conduct a comprehensive statistical analysis to rigorously assess the model's performance and reliability. Sections \ref{sec11} and \ref{sec12} present the results of our study and draw conclusions based on our findings, summarizing the contributions and implications of this research.\\\\\
\section{Cosmological equations in $f(Q,T)$ gravity}\label{sec2}
From the modification in Einstein-Hilbert action of general relativity by introducing a function of two scalar invariants, $Q$ and $T$, which are constructed from the non-metricity and the trace of the energy-momentum tensor. The action in $f(Q,T)$ is given by \cite{21}.
\begin{equation}
S=\int \left[\frac{1}{16\pi}{f(Q, T)}+\mathcal{L}_M\right]\sqrt{-g}\,d^{4}x  
\label{action}
\end{equation}
where, $g\equiv det(g_{\mu \nu})$, and $\mathcal{L}_M$ is Lagrangian density.\\
In Riemannian geometry, the metric tensor is always symmetric. However, for our research, we take the $f(Q,T)$ gravity, in which, we can take the non-symmetric part of the metric tensor called non-metricity. Which can be defined as 
\begin{equation}
Q\equiv -g^{\mu \nu }(L_{\beta \mu }^{\alpha }L_{\nu \alpha }^{\beta}-L_{\beta \alpha }^{\alpha }L_{\mu \nu }^{\beta })
\end{equation}
where 
\begin{equation}
    L_{\beta \gamma }^{\alpha }\equiv -\frac{1}{2}g^{\alpha \lambda }({\nabla}_{\gamma}g_{\beta \lambda}+{\nabla }_{\beta}g_{\lambda \gamma}-{\nabla }_{\lambda}g_{\beta \gamma})
\end{equation}
\newline
The trace of nonmetricity tensor is expressed as, 
\begin{equation}
Q_{\alpha }\equiv Q_{\alpha } \\
^{\mu } \\
_{\mu },\;{\tilde{Q}}_{\alpha }\equiv Q^{\mu } \\
_{\alpha \mu }
\end{equation}
The trace of the energy-momentum tensor and modification in the metric tensor
are respectively 
\begin{equation}
T_{\mu \nu }=-\frac{2}{\sqrt{-g}}\frac{\delta (\sqrt{-g}\mathcal{L}_{M})}{%
\delta g^{\mu \nu }}
\end{equation}%
\begin{equation}
\Theta _{\mu \nu }=g^{\alpha \beta }\frac{\delta T_{\alpha \beta }}{\delta
g^{\mu \nu }}
\end{equation}
Finding the variation of action of the field equation Eq. (\ref{action}) with respect to metric tensors.

\begin{widetext}
\begin{equation}
        8\pi T_{\mu \nu}=-\frac{2}{\sqrt{-g}}\nabla_\alpha ({f}_Q \sqrt{-g}P^{\alpha}_{\mu\nu}-\frac{1}{2}{f}g_{\mu\nu}+{f}_T (T_{\mu\nu}+\Theta_{\mu\nu})\\-{f}_Q (P_{\mu\alpha\beta} Q_{\nu}^{\alpha\beta}-2Q^{\alpha\beta_{\mu}}P_{\alpha\beta\nu}))
\end{equation}
Where super-momentum 
\begin{equation}
        P^{\alpha}_{\mu\nu}\equiv\frac{1}{4} \left[-Q^{\alpha}_{\mu\nu}+2Q_{(\mu} \,^{\alpha} \,_{\nu)} +Q^{\alpha} g_{\mu \nu}- \tilde{Q}^{\alpha} g_{\mu \nu}- \delta^{\alpha}\,_{(\mu}Q_{\nu)} \right]=-\frac{1}{2} L^{\alpha}_{\mu\nu}+\frac{1}{4}\left(Q^{\alpha}-\tilde{Q}^{\alpha} \right)g_{\mu\nu}-\frac{1}{4}\delta^{\alpha}\,_{(\mu}Q_{\nu )}
\end{equation}
\end{widetext}
Taking the FLRW metric as follows, 
\begin{equation}
ds^2=-N(t)^2dt^2+a(t)^2(dx^2+dy^2+dz^2),
\end{equation}
where N(t) is the lapse function and a(t) is the scale factor. Hence, $Q=6H^2/N^2$. We assume the value of N(t) = 1, for a standard case. Hence, $Q={6H^2}$.\\\\
To find the generalized Friedmann equations, assuming the matter content as the perfect fluid with the energy-momentum tensor $T_{\nu }^{\mu}=diag(-\rho,p,p,p)$. The tensor $\Theta _{\nu }^{\mu }$ becomes, 
\begin{equation}
\Theta _{\nu }^{\mu }=\delta _{\nu }^{\mu }p-2T_{\nu }^{\mu }=diag(2\rho+p,-p,-p,-p)
\end{equation}
For simplicity, taking $F\equiv {f}_{Q}= dF/dt$ and $8\pi \tilde{G}\equiv {f}_{T}=dF/dt$ the Friedmann equations we derived as
follows, 
\begin{equation}
8\pi \rho =\frac{{f}}{2}-6FH^{2}-\frac{2\tilde{G}}{1+\tilde{G}}(\dot{F}H+F\dot{H}) \label{f1}
\end{equation}
\begin{equation}
8\pi p=-\frac{{f}}{2}+6FH^{2}+2(\dot{F}H+F\dot{H})  \label{f2}
\end{equation}
From Eqs \eqref{f1} and \eqref{f2} , modified Einstein's field equations are derived 
\begin{equation}
3H^{2}=8\pi \rho _{eff}=\frac{{f}}{4F}-\frac{4\pi }{F}[(1+\bar{G})\rho +\bar{G}p]  \label{efe1}
\end{equation}
\begin{multline}
{2\dot{H}+3H^{2}}=-8\pi p_{eff}=\frac{{f}}{4F}-\frac{2\dot{F}H}{F} \\
+\frac{4\pi }{F}\left[ (1+\bar{G})\rho +(2+\bar{G})p\right]  \label{efe2}
\end{multline}
From Eqs. \eqref{efe1} and \eqref{efe2} and the derivative of Eq. \eqref{efe1} we derive the continuity equation 
\begin{equation}\label{eq15}
    \dot{\rho}_{eff}+3H\left( {\rho }_{eff}+{p}_{eff}\right) = 0
\end{equation}
Although there are several forms of the function $f(Q,T)$ is considered in the literature \cite{xu2020weyl}, we here only confined to the linear and additive form of $f(Q,T)$ function \cite{21,arora2020energy} in the form
\begin{equation}
f(Q,T)=\mu Q+\nu T  \label{model}
\end{equation}
where, $\mu $ and $\nu $ are the non zero model constants. Hence the first derivatives ${f}_{Q}=\mu $ and $8\pi \tilde{G}={f}_{T}=\nu $. Solving the modified Friedmann equations and applying the barotropic equation of states $p=\omega \rho$ we can find the equation of state parameter as follows
\begin{equation}
    \omega = \frac{3H^2 (8\pi+\nu) + \Dot{H} (16\pi +3\nu)}{\nu\Dot{H}-3H^2 (8\pi + \nu)}
    \label{eq17}
\end{equation}
Hence the energy density equation turns out to be 
\begin{equation}
    \rho = \frac{-3H^2 \mu (8\pi+\nu)+\mu\nu\Dot{H}}{2(4\pi +\nu)(8\pi +\nu)}
    \label{eq18}
\end{equation}
To find the value of $\Dot{H}$ we use the relation $a_0 / a=1+z$ we can define a new relation between z and t. 
\begin{equation}
    \frac{d}{dt}=\frac{dz}{dt}\frac{d}{dz}=-(1+z)H(z) \frac{d}{dz}
\end{equation} 
normalizing the equation by taking the value of the scale factor as $a_0=a(t=t_0)=1$, $t_0$ refers the present time. Hence we can write the derivation Hubble parameter with respect to time in terms of red-shift as, 
\begin{equation}
    \Dot{H} = -(1+z)H(z) \frac{dH}{dz}
    \label{tzdot}
\end{equation}

\section{The Model}\label{sec3}

The modified field equations, denoted as Eqs. \eqref{f1} and \eqref{f2} are
two independent equations describing the dynamics of the model. These
fundamental equations intertwined with three unknown variables: $a$, $\rho $%
, and $p$. This implies that, in order to completely characterize and solve
this system, we are in need of an additional equation. To address this,
researchers in the field have adopt a widely accepted approach \textit{the
model independent way} study of model, often considers a scheme of
parameterization of a cosmological parameter. In contemporary scientific
literature, this approach is gaining prominence, as it is capable of solving
the field equations in a model independent way that do not affect the
background physics but provide a solution in simple way. It involves making
an initial assumption regarding the scale factor, which can be done either
directly or by expressing it in terms of cosmological parameters and their
time derivatives. This strategic choice not only simplifies the equations
but also allows for a more comprehensive exploration of the model's
dynamics. To aid researchers in this endeavor, a wealth of parametrizations
for various cosmological parameters has been meticulously compiled and
organized in references \cite{29} and \cite{30}. These compilations serve as
invaluable resources, facilitating the selection of appropriate
parameterization schemes based on the specific characteristics and goals of
a given study. There are a few intriguing models of dark energy and modified
gravity based on various parametrization schemes of some geometrical
parameters \cite{33,34,35,36,37}. One noteworthy aspect of this concept is
its ability to reconstruct the cosmic history and also the fate, while
offering solutions to certain problems of standard cosmology.\newline
\newline
\qquad In our own investigation, we opt to employ the parameterization of
the deceleration parameter to effectively close the aforementioned system of
equations. This choice aligns with established practices, streamlining our
analytical framework while enabling us to delve deeply into the dynamics of
our model. We employ here a scheme of parametrization of higher order
time-dependent deceleration parameter ($q(t)$) in the form, 
\begin{equation}
q(t)=-1+\frac{m}{n}-\frac{4}{n}t^{3},  \label{dec}
\end{equation}%
where $m>0$ and $n>0$ are two arbitrary constants. Similar form is also
considered in the reference \cite{sofuouglu2023observational} in the context of $f(R,T)$
gravity. By using the definition of the deceleration parameter in the form
of Hubble parameter, $q=\frac{d}{dt}\left( \frac{1}{H}\right) -1$, we may
obtain an explicit form of Hubble function as 
\begin{equation}
H(t)=\frac{n}{t\left( m-t^{3}\right) }  \label{hub}
\end{equation}%
Note that this explicit form of the Hubble parameter can also be interpreted
as a particular form of the parametrization given in Ref. \cite{29} as a
special case. Using the definition of the Hubble parameter in terms of scale
factor will give the explicit form of scale factor as,
\begin{equation}
a=\beta \left( \frac{t^{3}}{m-t^{3}}\right) ^{\frac{n}{3m}},  \label{sf}
\end{equation}%
The deceleration parameter plays a crucial role in characterizing the
dynamics of the universe's expansion. It provides insights into whether the
expansion is slowing down or accelerating. Specifically, when the
deceleration parameter is greater than zero ($q>0$), it signifies a phase of
decelerating expansion. Conversely, when the deceleration parameter is less
than zero ($q<0$), it indicates an accelerating expansion phase. A
particularly interesting point occurs when the deceleration parameter equals
zero ($q=0$), marking a significant phase transition in the expansion of the
universe. This phase transition at $q=0$ occurs at a specific time, denoted
as $t_{tr}$, which can be calculated in this considered case as, $t_{tr}=%
\sqrt[3]{\frac{m-n}{4}}$. This implies, $m$ must be greater than $n$. This
moment represents a critical juncture in the universe's evolution, where the
expansion dynamics shift fundamentally. We can observe in Eqs. \eqref{hub}
and \eqref{sf}, for $t=m^{1/3}$, $H\longrightarrow \infty $ and $%
a\longrightarrow \infty $, implying the existence of a big rip singularity
in this model, which is anticipated to happen in the near future, precisely
at $t=t_{end}=m^{1/3}$. The universe's expansion reaches a point of extreme
instability and divergence, characterized by this singularity.
\section{Characterizing the Model through Dynamical Variables}\label{sec4}

This section is dedicated to providing a comprehensive analysis of the
physical behavior and properties inherent in our model. Within this context,
we aim to elucidate the model's physical dynamics by examining key
parameters, including energy density ($\rho $), pressure ($p$), and the
equation of state parameter ($\omega $). These parameters are essential in
understanding the behavior of our model and the physical processes it
encapsulates. The temporal or redshift evolution of these parameters serves
as a window into the dynamic nature of our model, shedding light on various
aspects of the Universe's evolution, particularly during its late stages and
with regard to stability considerations.\\\\
Our discussion in this section serves as the bedrock upon which a deeper
comprehension of our model is built. It lays the essential groundwork for
subsequent analyses and discussions. To initiate our exploration, we
leverage the field equations, denoted as Eqs. \eqref{f1} and \eqref{f2}.
These equations provide us with a means to express the energy density ($\rho 
$) and pressure ($p$) in specific mathematical forms. This mathematical
representation is pivotal in uncovering the intricate details of our model's
physical behavior and will serve as the basis for our further investigations
and interpretations. The explicit expressions for $\rho $ and $p$ using Eqs. %
\eqref{hub} and \eqref{sf} are found in this case as;

\begin{equation}
\rho (t)=\frac{-n\mu \lbrack \nu (m-4t^{3})+3n(8\pi +\nu )]}{%
2(-mt+t^{4})^{2}(4\pi +\nu )(8\pi +\nu )}  \label{rho1}
\end{equation}%
and 
\begin{equation}
p(t)=\frac{n\mu \lbrack 8\pi (-2m+3n+8t^{3})+3\nu (-m+n+4t^{3})]}{%
2(-mt+t^{4})^{2}(4\pi +\nu )(8\pi +\nu )}  \label{p1}
\end{equation}
The concept of the equation of state parameter denoted as $\omega $, holds
significant importance in the realm of cosmology. It serves as a fundamental
tool for characterizing the relationship between pressure ($p$) and energy
density ($\rho $) within a given system. This parameter is defined as the
ratio of pressure to energy density, and it plays a crucial role in
describing the thermodynamic behavior of fluids within the Universe. The
value of $\omega $ offers critical insights into how a substance or
component reacts to variations in volume or temperature. These insights, in
turn, have profound implications for the overall dynamics and fate of the
Universe. In this context, we have already derived expressions for both
energy density and pressure specific to our model.\\\\
Now, by employing Eqs. {\eqref{rho1} and \eqref{p1}}, we can deduce the
expression for the equation of state parameter $\omega $, as follows:
\begin{equation}
\omega (t)=\frac{8\pi (2m-3n-8t^{3})+3\nu (m-n-4t^{3})}{\nu
(m-4t^{3})+3n(8\pi +\nu )}  \label{eos}
\end{equation}

\section{Deriving Cosmological Parameters in terms of Redshift}\label{sec5}
To effectively constrain our model parameters, we need to express all the obtained cosmological variables in terms of the redshift
parameter, denoted as $z$. To achieve this, we utilize the relationship
between the redshift ($z$) and the scale factor ($a$), given by 
\begin{equation}
1+z=\frac{a_{0}}{a},
\end{equation}%
{where }$a_{0}${\ is the present value of the scale factor and generally
normalized to be }$a_{0}=1$. {The} $t-z$ relationship in our case can be
established as; 
\begin{equation}
t=\sqrt[3]{k_{1}}\left\{ 1+[\zeta (1+z)]^{3\frac{m}{n}}\right\} ^{-\frac{1}{3%
}}  \label{tz}
\end{equation}%
Now, the Hubble parameter in terms of redshift $z$ for the Model is,
\begin{equation}
H(z)=H_{0}\left(1+\zeta^{3 \eta}\right)^{-\frac{4}{3}}(1+z)^{-3 \eta}\left\{1+[\zeta(1+z)]^{3 \eta}\right\}^{\frac{4}{3}}, 
\end{equation}
where $\frac{m}{n} = \eta$. Now, the deceleration parameter, energy density, pressure, and the equation of state parameter can be rewritten in terms of redshift $z$ as follows;
\begin{equation}
q(z)=-1+\eta-\frac{4\eta}{\left\{ 1+[\zeta (1+z)]^{3\eta }\right\} },
\label{qz}
\end{equation}
\begin{widetext}

\begin{equation}
p(z) = -\frac{{\mu \, H_0^2 \, (1 + (\zeta \, (1 + z))^{3\eta})^{5/3} \, (16\pi + 3\nu) \, \eta \, (-3 + (\zeta \, (1 + z))^{3\eta}) - 3 \, (8\pi + \nu) \, (1 + (\zeta \, (1 + z))^{3\eta})}}{{2 \, (4\pi + \nu) \, (8\pi + \nu) \, (1 + \zeta^2 (1 + z)^{3\eta})^{8/3}}}
\end{equation}

\begin{equation}
\rho (z) = - \frac{\left( (1 + z)^{-6\eta} \mu H_0^2 \left( 1 + \left( (1 + z) \zeta \right)^{3\eta} \right)^{5/3} \left( 3 \left( 8\pi + \nu - \nu \eta \right) + \left( 3 \left( 8\pi + \nu \right) + \nu \eta \right) \left( (1 + z) \zeta \right)^{3\eta} \right) \right)}{2 \left( 4\pi + \nu \right) \left( 8\pi + \nu \right) \left( 1 + \zeta^3  \eta^8 \right)^{1/3}}
\end{equation}
\begin{equation}
\omega (z)=\frac{{(16\pi + 3\nu)\eta \left( -3 + \left( (1 + z)\zeta \right)^{3\eta} \right) - 3(8\pi + \nu) \left( 1 + \left( (1 + z)\zeta \right)^{3\eta} \right)}}{{3\left( 8\pi + \nu - \nu\eta \right) + 3\left( 8\pi + \nu \right) + \nu\eta \left( (1 + z)\zeta \right)^{3\eta}}}
\end{equation}
\end{widetext}

To comprehensively explore the evolution of both geometric and physical parameters, as well as to rigorously assess the validity of our
derived model, it becomes imperative to acquire precise values for the model
parameters that play a pivotal role in our theoretical framework. Thus, we
perform the data analysis in the subsequent section to derive some precise
values of the model parameter. Through this data analysis, we aim to gain a
deeper understanding of the dynamics governing our cosmological model and
evaluate its consistency with observation too.

\section{Data Analysis} \label{sec6}
In this section, we conduct an extensive comparison between our proposed cosmological model and a wide range of available cosmological data. Our aim is to understand the fundamental characteristics of the model through various datasets, including the Cosmic Chronometers (CC), type Ia supernovae (SNIa), Gamma Ray Bursts (GRBs), Quasars (Q), Baryon Acoustic Oscillation (BAO) and Cosmic Microwave Background (CMB) This rigorous investigation aims to determine the optimal values of key model parameters, such as $\eta$, $\zeta$. These datasets together describe how our model behaves in the context of the $f(Q,T)$ gravity framework. Importantly, we also account for the present-day Hubble function, denoted as $H_{0}$, which plays a crucial role in shaping our results. To obtain best fit values our cosmological model we employ a robust Bayesian statistical methodology. This method relies on likelihood functions and a well-accepted technique called Markov Chain Monte Carlo (MCMC). Within this Bayesian framework, we construct a probabilistic assessment of how likely certain combinations of model parameters are based on actual observations. Our investigation reveals hidden aspects of our cosmological model, offering valuable insights into how it deeply connects with the observable universe.

\subsection{Methodology}\label{methodology}
Constraining the Hubble function through observational data involves a process known as parameter estimation or model fitting. In our case, our goal is to determine the optimal values of key model parameters, using various datasets , including Cosmic Chronometers (CC), Type Ia Supernovae (SNIa), Gamma-Ray Bursts (GRBs), Quasars (Q), Baryon Acoustic Oscillations (BAO), and the Cosmic Microwave Background (CMB). The initial step is to establish a likelihood function that measures the agreement between our model predictions and the observed data.
\begin{equation}
\mathcal{L}(\theta) = \exp\left(-\frac{1}{2} \sum_{i=1}^{N} \frac{(O_i - M_i(\theta))^2}{\sigma_i^2}\right)
\end{equation}
Where \(O_i\) is the observed data point for the ith data point, \(M_i(\theta)\) is the model prediction for the ith data point based on the parameters \(\theta\), and \(\sigma_i\) represents the uncertainty associated with the observed data point. Next, we perform Bayesian parameter estimation \cite{trotta2017bayesian}. This requires defining prior distributions for each parameters we want to constrain, which should encapsulate any existing knowledge or constraints. If strong prior information is lacking, relatively flat priors can be used. The posterior distribution, proportional to the likelihood function multiplied by the prior distribution, is then computed as:
\begin{equation}
P(\theta | D) \propto \mathcal{L}(\theta) \times \pi(\theta)    
\end{equation}
In the subsequent steps, Markov Chain Monte Carlo (MCMC) is employed to explore the posterior distribution and derive parameter constraints \cite{akeret2013cosmohammer}. This widely employed sampling technique generates an extensive set of samples from the posterior distribution, allowing for the extraction of key statistical measures. These measures include mean, median, standard deviation, and credible intervals for each parameter, providing optimal parameter values and their associated uncertainties. Subsequently, we rigorously evaluate our model's performance against the Cosmic Chronometers (CC) and type Ia supernovae (SNIa) datasets. Visual comparisons and quantitative metrics, such as chi-squared values or the Akaike Information Criterion (AIC), are used to assess the model's fit. The ensuing discussions delve into the implications of the derived parameter constraints, emphasizing the model's alignment with observational data. We explore potential physical interpretations and ramifications, enriching our understanding of the underlying cosmological dynamics.

\subsection{Data Discription}
\subsubsection{Cosmic Chronometers}
In our analysis, we utilize thirty-one data points acquired through the cosmic chronometers (CC) technique for the determination of the Hubble parameter. This approach allows us to directly extract information about the Hubble function at various redshifts, extending up to $z \lesssim 2$. The selection of CC data is motivated by its reliability, as it primarily involves measurements of the age difference between two passively evolving galaxies that originated at the same time but have a slight separation in redshift. This technique enables us to compute $\Delta z / \Delta t$, making CC data preferable to methods based on absolute age determinations for galaxies \cite{72CC}. Our chosen CC data points were sourced from independent \cite{73CC,74CC,75CC,76CC,77CC,78CC,79CC}. Importantly, these references are not influenced by the Cepheid distance scale or any specific cosmological model. Nevertheless, it's worth noting that they do depend on the modeling of stellar ages, which is established using robust stellar population synthesis techniques (for more details, see \cite{75CC,77CC,80CC,82CC,81CC,83CC} for analyses related to CC systematics). We evaluate the goodness of fit using the $\chi_{CC}^{2}$ estimator, which is expressed as follows:

\begin{equation}
\chi_{CC}^{2}(\Theta)=\sum_{i=1}^{31} \frac{\left(H\left(z_{i}, \Theta\right)-H_{\mathrm{obs}}\left(z_{i}\right)\right)^{2}}{\sigma_{H}^{2}\left(z_{i}\right)},
\end{equation}
Here, $H\left(z_{i}, \Theta\right)$ represents the theoretical Hubble parameter values at redshift $z_{i}$ with model parameters denoted as $\Theta$. The observational data for the Hubble parameter at $z_{i}$ is given by $H_{\mathrm{obs}}\left(z_{i}\right)$, with an associated observational error of $\sigma_{H}\left(z_{i}\right)$.

\subsubsection{type Ia supernovae (SNIa)}
Over the years, a multitude of supernova datasets has been established, including references such as \cite{Pan1,Pan2,Pan3,Pan4,Pan5}. Recently, a refreshed version of the Pantheon dataset, referred to as Pantheon+, has been introduced \cite{pantheon+}. This updated compilation comprises 1701 data points of type Ia supernovae (SNIa), spanning the redshift interval $0.001<z<2.3$. SNIa observations have played a pivotal role in unveiling the phenomenon of the universe's accelerating expansion. These observations serve as crucial tools for investigating the nature of the driving component behind this expansion, owing to SNIa's status as luminous astrophysical objects. These objects, often treated as standard candles, enable the measurement of relative distances based on their intrinsic brightness. The Pantheon+ dataset stands as a valuable resource, offering insights into the accelerating universe's characteristics. The chi-square statistic serves as a fundamental tool for comparing theoretical models with observational data. In the context of the Pantheon+ dataset, chi-square values are computed using the subsequent equation

\begin{equation}
\chi_{\text{Pantheon+}}^2 = \vec{D}^T \cdot \mathbf{C}_{\text{Pantheon+}}^{-1} \cdot \vec{D}
\end{equation}
Here, $\vec{D}$ represents the difference between the observed apparent magnitudes $m_{Bi}$ of SNIa and the expected magnitudes given by the cosmological model. $M$ represents the absolute magnitude of SNIa, and $\mu_{\text{model}}$ is the corresponding distance modulus predicted by the assumed cosmological model. The term $\mathbf{C}_{\text{Pantheon+}}$ denotes the covariance matrix provided with the Pantheon+ data, which includes both statistical and systematic uncertainties. The distance modulus is a measure of the distance to an object, defined as:
\begin{equation}
\mu_{\text{model}}(z_i) = 5\log_{10}\left(\frac{D_L(z_i)}{(H_0/c) \text{ Mpc}}\right) + 25
\end{equation}
Here, $D_L(z)$ represents the luminosity distance, which is calculated for a flat homogeneous and isotropic FLRW universe as:
\begin{equation}
D_L(z) = (1+z)H_0\int_{0}^{z}\frac{dz'}{H(z')}
\end{equation}
The Pantheon+ dataset differs from the previous Pantheon sample as it breaks the degeneracy between the absolute magnitude $M$ and the Hubble constant $H_0$. This is achieved by rewriting the vector $\vec{D}$ in terms of the distance moduli of SNIa in the Cepheid hosts. The distance moduli in the Cepheid hosts, denoted as $\mu_i^{\text{Ceph}}$, are measured independently using Cepheid calibrators. This allows for the independent constraint of the absolute magnitude $M$. The modified vector $\vec{D'}$ is defined as:
\begin{equation}
\vec{D'}_i = \begin{cases}
m_{Bi} - M - \mu_i^{\text{Ceph}} & \text{if } i \text{ is in Cepheid hosts} \\
m_{Bi} - M - \mu_{\text{model}}(z_i) & \text{otherwise}
\end{cases}
\end{equation}
With this modification, the chi-square equation for the Pantheon+ dataset can be rewritten as:
\begin{equation}
\chi_{\text{SN}}^2 = \vec{D'}^T \cdot \mathbf{C}_{\text{Pantheon+}}^{-1} \cdot \vec{D'}
\end{equation}
This revised formulation allows for improved constraints on the absolute magnitude $M$ and the cosmological parameters.\\\\
We've also expanded our investigation to include a subset of 162 Gamma Ray Bursts (GRBs) \cite{GRB}, spanning a redshift range of $1.44<z<8.1$. In this context, we define the $\chi^2$ function as:

\begin{equation}
    \chi_{\text{GRB}}^2(\phi_{\text{g}}^\nu) = \mu_{\text{g}} \mathbf{C}_{\text{g,cov}}^{-1} \mu_{\text{g}}^T,
\end{equation}
Here, $\mu_{\text{g}}$ denotes the vector encapsulating the differences between the observed and theoretical distance moduli for each individual GRB. Similarly, for our examination of 24 compact radio quasar observations \cite{quasers}, spanning redshifts in the range of $0.46\leq z\leq 2.76$, we establish the $\chi^2$ function as:

\begin{equation}
    \chi_{\text{Q}}^2(\phi_{\text{q}}^\nu) = \mu_{\text{q}} \mathbf{C}_{\text{q,cov}}^{-1} \mu_{\text{q}}^T,
\end{equation}
In this context, $\mu_{\text{q}}$ represents the vector capturing the disparities between the observed and theoretical distance moduli for each quasar.

\subsubsection{Baryon Acoustic Oscillations}
To study Baryon Acoustic Oscillations (BAO), we utilize a dataset consisting of 333 measurements \cite{baonew1,baonew2,bao3,bao4,bao5,bao6,bao7,bao8,bao9,bao10,bao11,bao12}. However, considering the potential error arising from data correlations, we select a smaller dataset of 17 BAO measurements for our analysis (please see table 1 of this work \cite{benisty2021testing} ). This selection helps to reduce errors and improve the accuracy of our results. One of the key measurements obtained from BAO studies in the transverse direction is the quantity $D_H(z)/r_d$, where $D_H(z)$ represents the comoving angular diameter distance. It is related to the following expression \cite{bao13,bao14}:

\begin{equation}
D_M = \frac{c}{H_0} S_k\left(\int_0^z \frac{d z^{\prime}}{E\left(z^{\prime}\right)}\right),
\end{equation}

where $S_k(x)$ is defined as:

\begin{equation}
S_k(x) = \begin{cases}\frac{1}{\sqrt{\Omega_k}} \sinh \left(\sqrt{\Omega_k} x\right) & \text { if } \quad \Omega_k>0 \\ x & \text { if } \quad \Omega_k=0 \\ \frac{1}{\sqrt{-\Omega_k}} \sin \left(\sqrt{-\Omega_k} x\right) & \text { if } \quad \Omega_k<0 .\end{cases}
\end{equation}

Additionally, we consider the angular diameter distance $D_A = D_M / (1+z)$ and the quantity $D_V(z)/r_d$. The latter is a combination of the coordinates of the BAO peak and $r_d$, representing the sound horizon at the drag epoch. Furthermore, we can directly obtain "line-of-sight" or "radial" observations from the Hubble parameter using the expression:
\begin{equation}
D_V(z) \equiv \left[z D_H(z) D_M^2(z)\right]^{1/3}.
\end{equation}
By studying these BAO measurements, we gain insights into the cosmological properties and evolution of the universe, while minimizing potential errors and considering relevant distance measures and observational parameters.\\\\
\subsubsection{Cosmic Microwave Background}\label{CMB}
The CMB distant prior measurements are taken \cite{chen2019distance}. The distance priors
offer useful details about the CMB power spectrum in two ways: the acoustic
scale $l_{A}$ characterizes the CMB temperature power spectrum in the
transverse direction, causing the peak spacing to vary, and the "shift
parameter" $R$ influences the CMB temperature spectrum along the
line-of-sight direction, affecting the peak heights, which are defined as
follows:
\begin{equation}
l_A=(1+z_d) \frac{\pi D_A(z)}{r_s},
\end{equation}
\begin{equation}
\quad R(z)=\frac{\sqrt{\Omega_m} H_0}{c}(1+z_d) D_A(z)
\end{equation}
The observables that \cite{chen2019distance} reports are:$R_{z}=1.7502\pm 0.0046,\quad
l_{A}=301.471\pm 0.09,\quad n_{s}=0.9649\pm 0.0043$ and $r_{s}$ is an
independent parameter, with an associated covariance matrix. (see table I in 
\cite{chen2019distance}). The points represent the inflationary observables as well as the CMB epoch expansion rate. In addition to the CMB points, we also take into account other data from the late Universe. The result is a successful test
of the model in relation to the data.


\begin{figure}[H]
\centering
\includegraphics[scale=0.5]{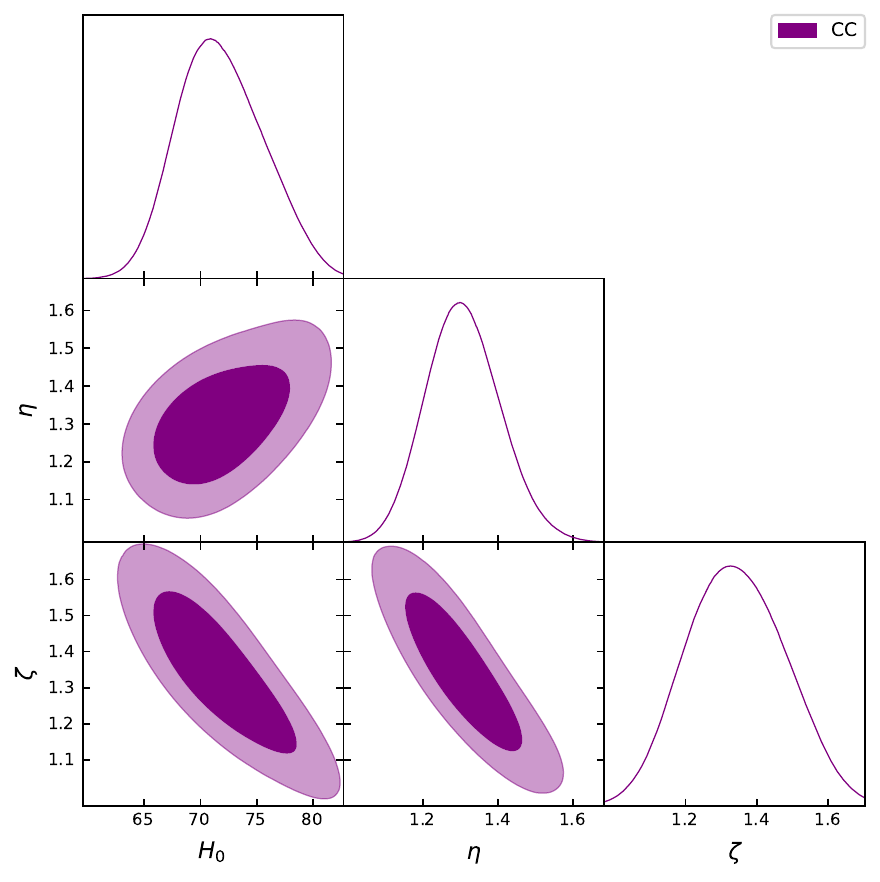}
\caption{The above figure shows the MCMC confidence contours at 1$\sigma$ and 2$\sigma$  obtained from the CC dataset.}\label{fig1}
\end{figure}

\begin{figure}[H]
\centering
\includegraphics[scale=0.5]{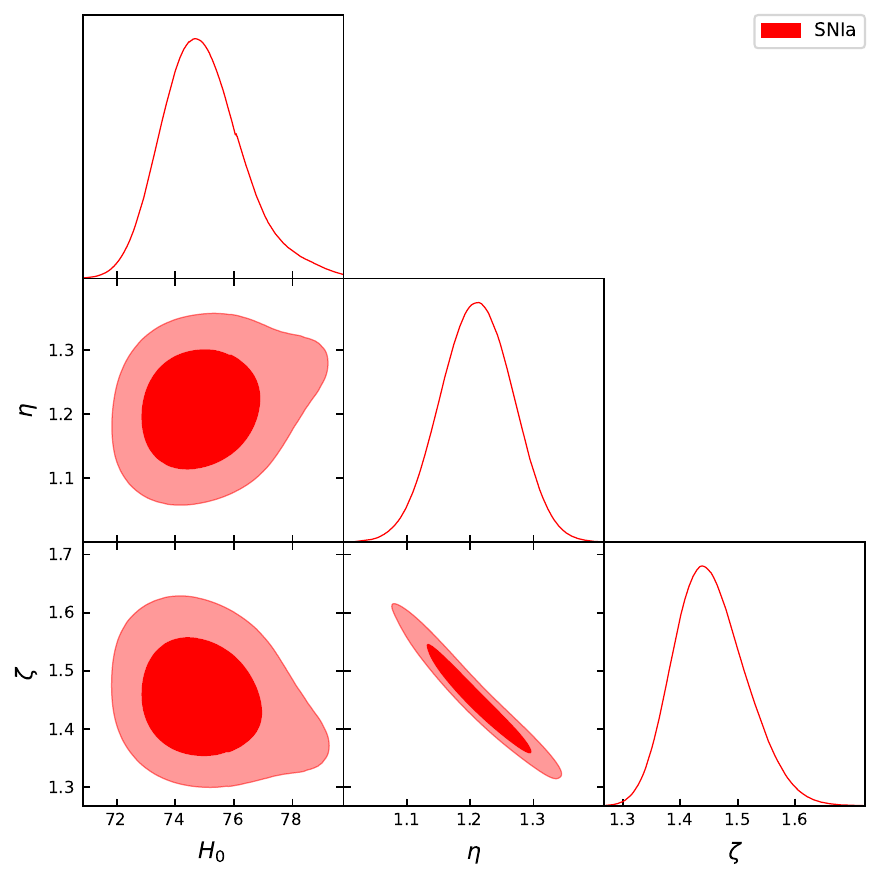}
\caption{The above figure shows the MCMC confidence contours at 1$\sigma$ and 2$\sigma$  obtained from the SNIa dataset.}\label{fig2}
\end{figure}

\begin{figure}[H]
\centering
\includegraphics[scale=0.5]{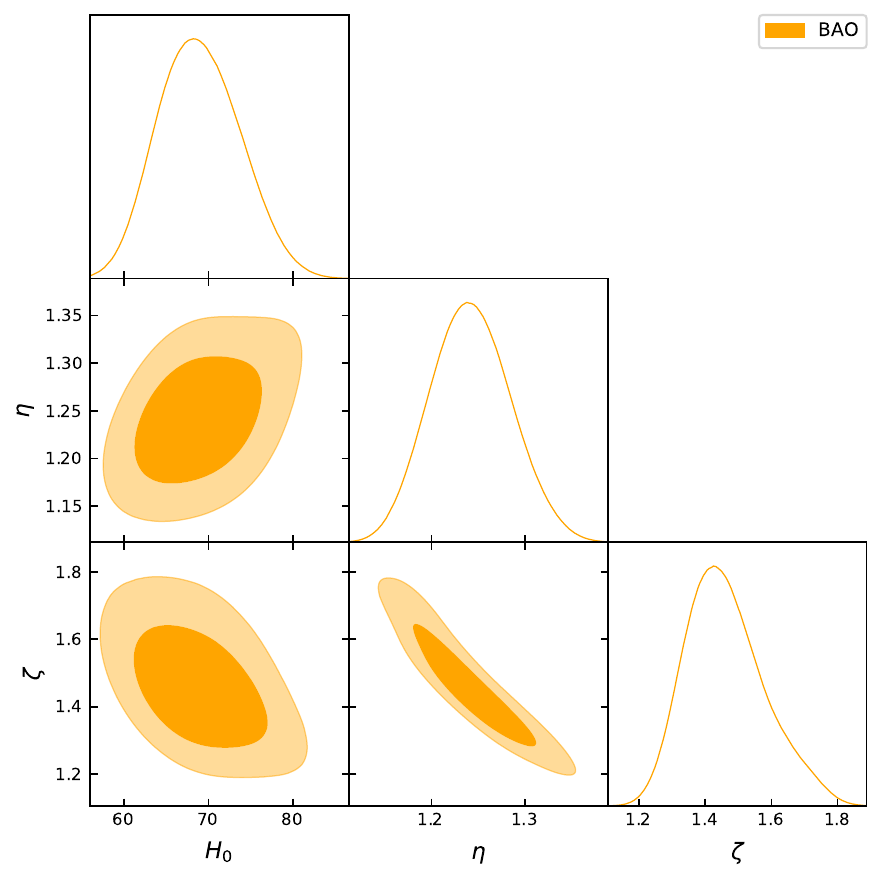}
\caption{The above figure shows the MCMC confidence contours at 1$\sigma$ and 2$\sigma$  obtained from BAO dataset.}\label{fig3}
\end{figure}

\begin{figure}[H]
\centering
\includegraphics[scale=0.5]{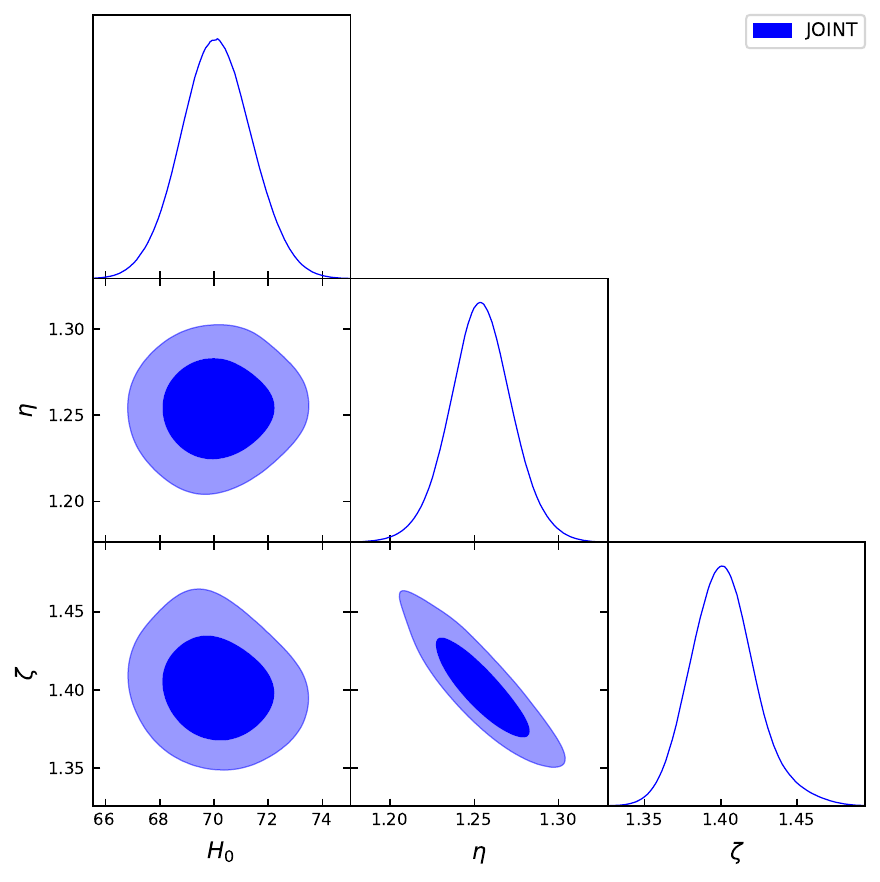}
\caption{The above figure shows the MCMC confidence contours at 1$\sigma$ and 2$\sigma$  obtained from the CC + SNIa + GRB + Q + BAO + CMB  dataset.}\label{fig4}
\end{figure}

\begin{figure}[H]
\centering
\includegraphics[scale=0.5]{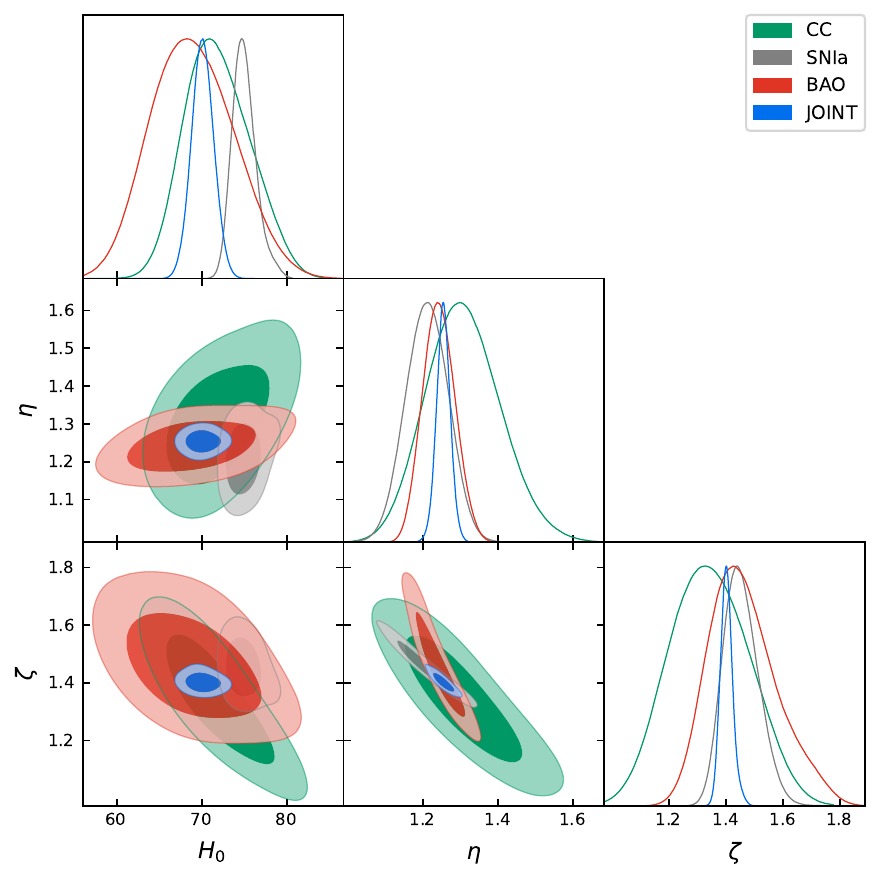}
\caption{The above figure shows the MCMC confidence contours at 1$\sigma$ and 2$\sigma$ with respect to All dataset}\label{fig5}
\end{figure}


\begin{widetext}.
\begin{table}[H]
\begin{center}
\begin{tabular}{c c c c c c c}
\hline
Model & Dataset & $H_{0}$ & $\eta$ & $\zeta$ & $q(z=0)$ & $j(z=0)$ \\[1ex] \hline \hline
$\Lambda$CDM Model & CC & $68.080661^{+2.121480}_{-2.121480}$ & --- & --- &$-0.52658^ {+0.070940}_{-0.095155}$& $1$ \\ 

& SNIa & $74.174328^{+ 1.267662}_{-1.267662}$ &---&---& $-0.56965^ {+0.019986}_{-0.040265}$ & $1$  \\ 

& BAO & $69.089209^{+4.396874}_{-4.396874}$& --- &---& $-0.60418^ {+0.014706}_{-0.027917}$ & $1$  \\ 

& Joint & $69.854848^{+1.259100}_{-1.259100}$ & --- & --- & $-0.59092^ {+0.018813}_{-0.035450}$ & $1$  \\ \hline
   
Model & CC &  $ 71.819162^{+ 3.670608}_{- 5.331030}$& $1.306004^{+0.092016}_{-0.171492}$ & $ 1.341120^{+0.142478}_{-0.216258}$ & $-1.30926^ {+0.144385}_{-0.218206}$ & $5.1455^ {+0.096127}_{-0.178301}$ \\ 

& SNIa &$74.945259^{+ 1.405723}_{-2.176923}$&$1.210586^{+0.052141}_{-0.099643}$& $1.451716^{+0.065137}_{-0.088348}$& $-1.17116^ {+0.143365}_{-0.212348}$ & $6.24952^ {+0.092561}_{-0.176606}$  \\ 
 
& BAO & $68.882349^{+ 4.806346}_{- 7.794739}$ &$1.241916^{+0.041110}_{-0.077956}$& $ 1.459481^{+0.116204}_{-0.202248}$ & $-1.15407^ {+0.006608}_{-0.009882}$& $6.70014^ {+0.028481}_{-0.072066}$ \\ 
 
& Joint &$70.090386^{+ 1.244084}_{- 2.612879}$  &$1.253896^{+0.017497}_{-0.037139}$&  $ 1.401909^{+0.021304}_{-0.039846}$ & $-1.02231^ {+0.006802}_{-0.010784}$ & $7.44366^ {+0.029072}_{-0.076997}$  \\ \hline
\hline
\end{tabular}%
\end{center}
\caption{We present constraints on cosmological parameters for both the standard $\Lambda$CDM and proposed models at a 95\% confidence level (CL). These constraints are derived from the CC dataset, Type Ia supernovae (SNIa), and Baryon Acoustic Oscillations (BAO) datasets. Additionally, the 'JOINT' dataset combines the 'CC' dataset with data from Type Ia supernovae, gamma-ray bursts, quasars, BAO, and CMB distant prior measurements. The table includes values for $q(z=0)$, $j(z=0)$, and $\chi^2_{\text{min}}$ with respect to different datasets.}
\label{MCMC final Output}
\end{table}
\end{widetext}
The contour plots derived from various datasets, including the CC dataset, BAO dataset, SNIa dataset, and the combined dataset (CC + SNIa + GRB + Q + BAO + CMB), are presented in Figure \ref{fig1}, Fig \ref{fig2}, Fig \ref{fig3}, and Fig \ref{fig4}, respectively. To provide a comprehensive overview, we have overlaid all four figures from these diverse datasets in Fig \ref{fig5}. In Table \ref{MCMC final Output}, we have tabulated the best-fit values of the model parameters $\eta$ and $\zeta$, along with the present-day Hubble function $H_{0}$, complete with their corresponding error bars and $q(z=0)$, $j(z=0)$, and $\chi^2_{\text{min}}$ with respect to different datasets.
\section{Observational and theoretical comparisons of the Hubble Function and Distance Modulus Function}\label{sec7}
After obtaining the best-fit values for our cosmological model parameters, it's crucial to compare our model with the widely accepted $\Lambda$CDM model. The $\Lambda$CDM model has consistently aligned with various observational datasets and stands as a robust framework for understanding the Universe's evolution. This comparative analysis deepens our comprehension of differences between the two models, shedding light on the implications of these disparities in cosmology. By examining deviations between our model and the $\Lambda$CDM model, we pinpoint the specific features that distinguish our parametrized model, such as the Universe's dynamics. This exploration provides valuable insights into our model's strengths and limitations, enriching our understanding of the cosmos. This comparison with the $\Lambda$CDM model acts as a benchmark, allowing us to assess the goodness-of-fit to observational data and gauge the alignment between our parametrized model and the established $\Lambda$CDM framework.
\subsection{Comparison with the CC data points}
\begin{figure}[!htb]
   \begin{minipage}{0.49\textwidth}
     \centering
   \includegraphics[scale=0.43]{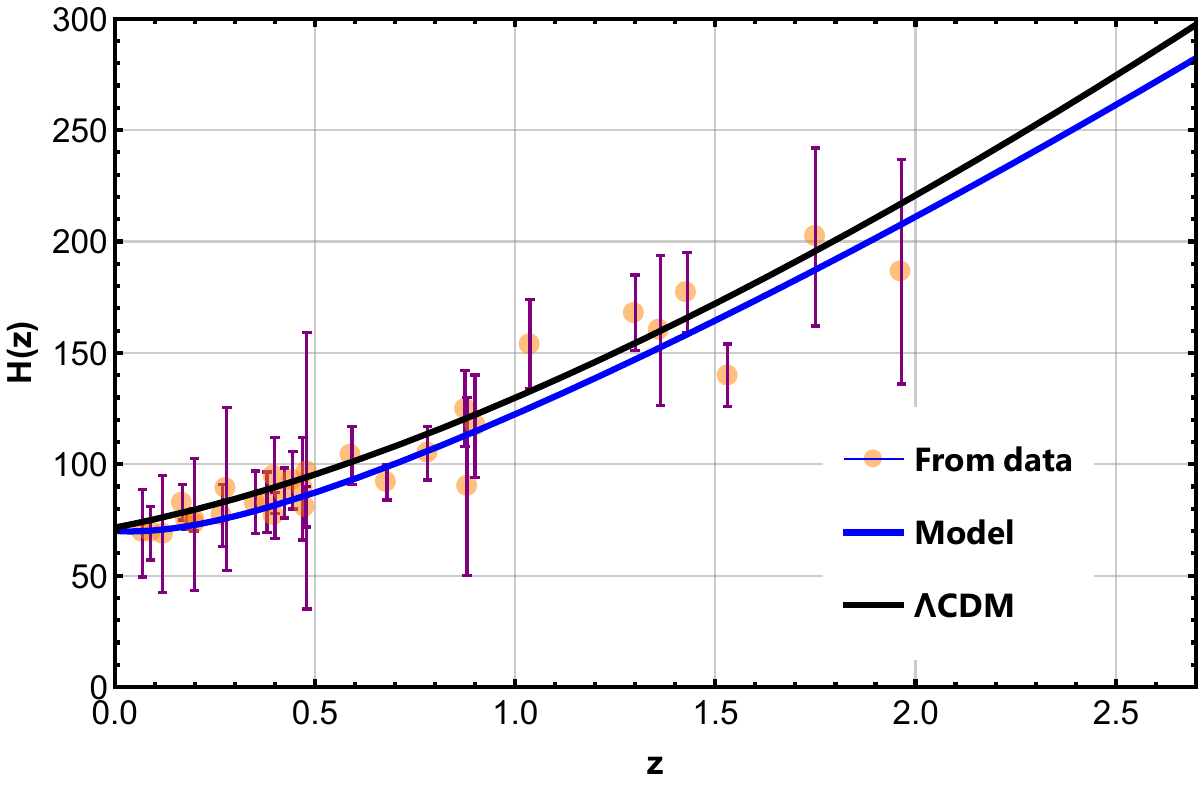}
\caption{Comparative analysis of our model ( Blue Line ) with 31 CC measurements ( Orange dots ) and $\Lambda$CDM model ( black line ).}\label{fig6}
   \end{minipage}
\end{figure}

We evaluate our model's compatibility with observational data by conducting a comparative analysis with the Cosmic Chronometers dataset, consisting of 31 data points represented by orange dots, each accompanied by error bars represented by purple dots. This comparison is illustrated in Fig~\ref{fig6}, where our model's predictions are depicted by the purple line. To provide a reference point, we also include the well-established $\Lambda$CDM model, represented by the black line, with cosmological parameters $\Omega_{\mathrm{m0}}=$ 0.3 and $\Omega_\Lambda =$ 0.7. The results of this analysis reveal a remarkable correspondence between our model's predictions and the observed data points, as indicated by their close alignment. This alignment underscores the ability of our model to effectively capture the inherent features and trends within the Cosmic Chronometers dataset. Consequently, our model demonstrates its capability to reproduce the expansion history of the universe, as inferred from the Hubble data.\\\\
\subsection{Comparison with the type Ia supernova dataset}
\begin{figure}[!htb]
   \begin{minipage}{0.49\textwidth}
     \centering
   \includegraphics[scale=0.35]{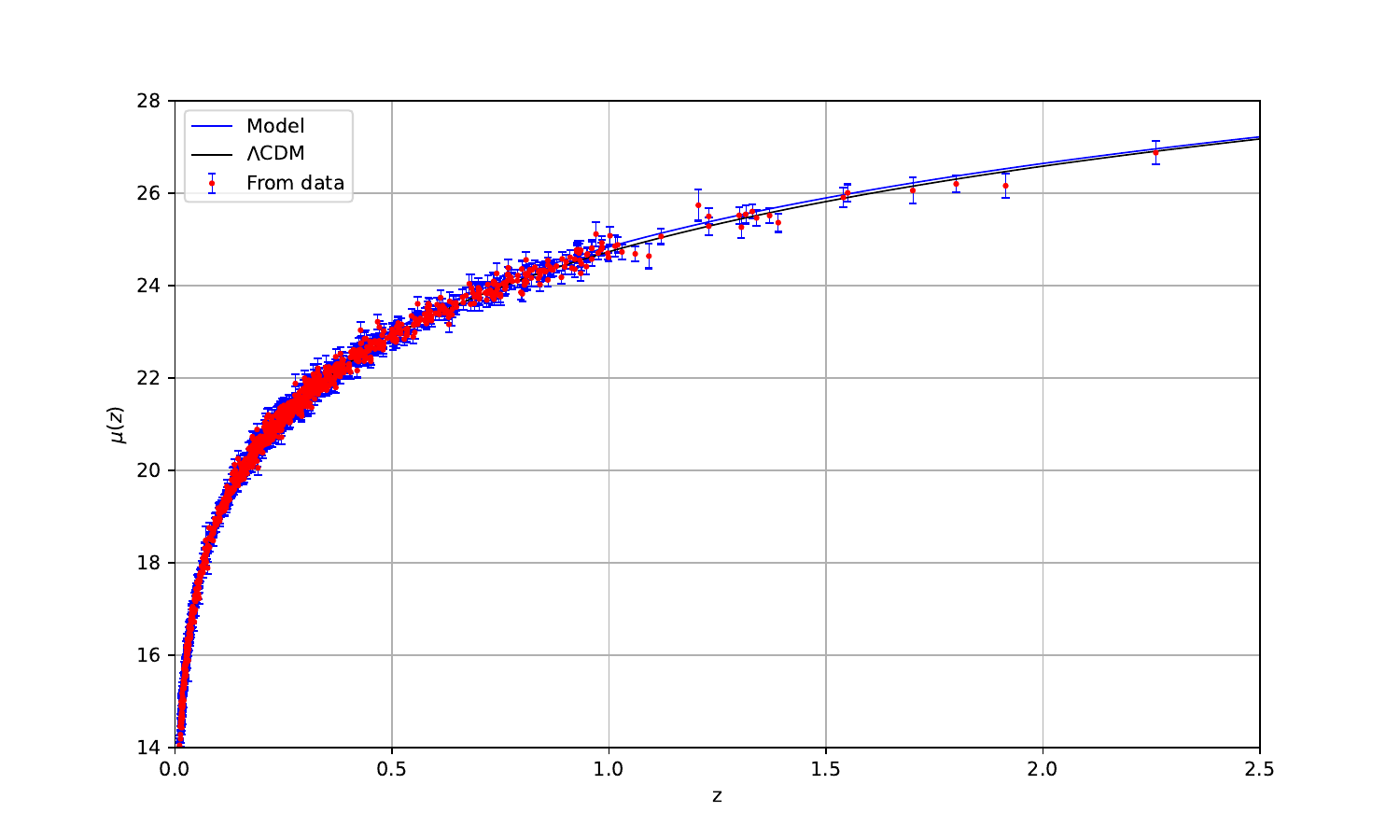}
\caption{Comparative analysis of our model ( Blue Line ) with 1701 type Ia supernova measurements ( Orange dots ) and $\Lambda$CDM model ( black line ).}\label{mu(z) Model 1}
   \end{minipage}
\end{figure}

In this analysis, we carefully studied the $\mu(z)$ distance modulus function for our Model alongside the data from type Ia supernovae, which includes a significant 1701 data points. We also compared our Model with standard $\Lambda$CDM Model. The results of this comparison are represented in Fig \ref{mu(z) Model 1}. The figures clearly illustrate that our model and the $\Lambda$CDM model exhibit a commendable level of agreement with the type Ia supernova dataset. This agreement indicates that these models effectively capture and replicate the observed distance measurements, signifying their consistency with empirical astronomical data. \\\\
\subsection{Relative difference between model and $\Lambda$CDM}
\begin{figure}[!htb]
   \begin{minipage}{0.49\textwidth}
     \centering
\includegraphics[scale=0.41]{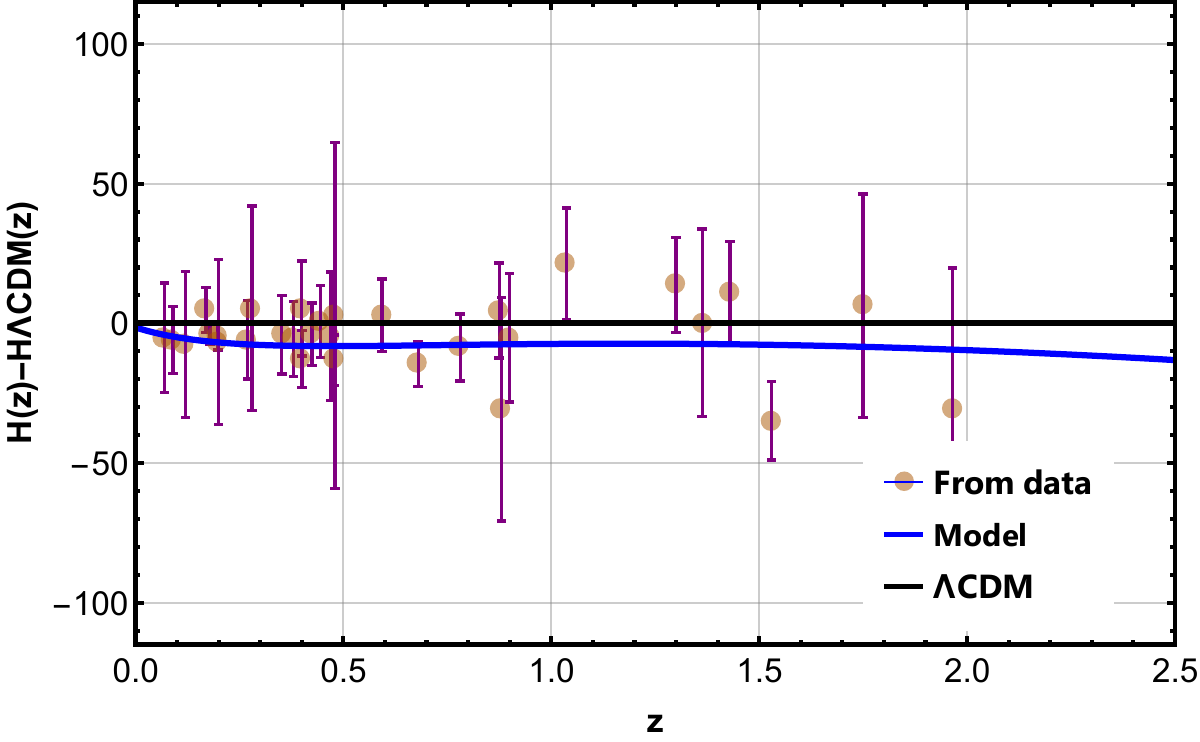}
\caption{Comparative analysis of our model ( Blue Line ) with 31 CC measurements ( Orange dots ) and $\Lambda$CDM model ( black line ).}\label{h(z)diff1}
   \end{minipage}
\end{figure}

The Figure \ref{h(z)diff1} illustrates the relative difference between our Model and the standard $\Lambda$CDM model. It is evident from the Figure that our Model exhibits distinct behavior from the typical $\Lambda$CDM model, particularly noticeable start form low redshift values ($z>0$). As the redshift increases, these disparities between our model and the $\Lambda$CDM Model become more pronounced.

\section{Cosmography Parameters}\label{sec8}
\subsection{The deceleration parameter}
The deceleration parameter \cite{visser2005cosmography}, denoted as "$q$," is a fundamental cosmological parameter used in the study of the expansion dynamics of the universe. It plays a crucial role in understanding the past and future evolution of the cosmos. Introduced by Edwin Hubble in the early 20th century, this parameter provides insights into whether the expansion of the universe is slowing down or accelerating. The deceleration parameter is defined in terms of the second derivative of the scale factor of the universe, which describes how the universe's size changes with time. Mathematically, it can be expressed as:
\begin{equation}
    q = -\frac{a\ddot{a}}{\dot{a}^2},
\end{equation}
where $q$ is the deceleration parameter. $a(t)$ represents the scale factor of the universe as a function of time. $\dot{a}$ represents the first derivative of the scale factor with respect to time. $\ddot{a}$ represents the second derivative of the scale factor with respect to time. The value of the deceleration parameter is indicative of the nature of the cosmic expansion:
\begin{enumerate}
    \item $q > 0$ (Decelerating Universe): If the deceleration parameter is positive, it implies that the expansion of the universe is slowing down over time. In the past, this was the prevailing belief when the gravitational attraction of matter was thought to dominate the cosmic dynamics.

    \item $q = 0$ (Critical Universe): A deceleration parameter of zero suggests that the expansion is proceeding at a constant rate. In this scenario, the universe's expansion is neither accelerating nor decelerating, often referred to as a "critical universe."

    \item $q < 0$ (Accelerating Universe): A negative value for the deceleration parameter signifies that the universe's expansion is accelerating. This phenomenon gained significant attention in the late 20th century when the discovery of dark energy provided a compelling explanation for this observed acceleration.
\end{enumerate}
In recent years, the study of the deceleration parameter has become increasingly important in cosmology, particularly in the context of understanding dark energy and the fate of the universe. It is a key parameter used in observational cosmology to probe the nature of the components of the universe, such as dark matter and dark energy, and to determine the overall geometry of the universe.
\begin{figure}[!htb]
   \begin{minipage}{0.49\textwidth}
     \centering
   \includegraphics[scale=0.43]{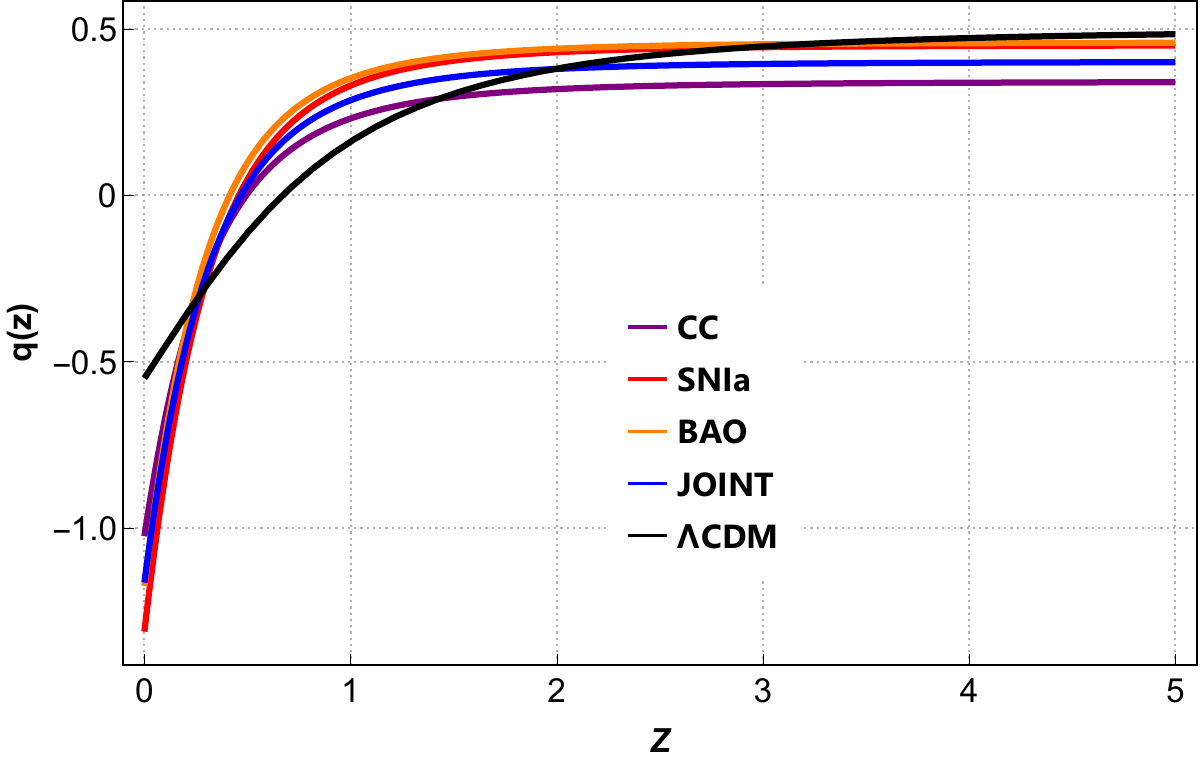}
\caption{Evolution of the deceleration parameter with respect to redshift across various datasets, depicted alongside the $\Lambda$CDM model represented by the black line, characterized by $\Omega_{\mathrm{m0}}=0.3$ and $\Omega_\Lambda=0.7$.}\label{fig_9}
   \end{minipage}
\end{figure}
\subsection{The jerk parameter}
In the realm of cosmology, understanding the dynamics of the universe's expansion is of paramount importance. While the Hubble constant and the deceleration parameter have been essential tools in characterizing this expansion, a more nuanced parameter known as the "jerk parameter" has emerged as a valuable addition to our cosmological toolkit. The jerk parameter, denoted as "j," provides a deeper level of insight into the cosmic acceleration, complementing the information offered by the deceleration parameter \cite{visser2004jerk}. The jerk parameter represents the third time derivative of the scale factor of the universe, building upon the concepts embodied by the Hubble parameter and the deceleration parameter. Mathematically, it can be expressed as:
\begin{equation}
    j = \frac{1}{a}\frac{d^3a}{d\tau^3}\left[\frac{1}{a}\frac{da}{d\tau}\right]^{-3} = q(2q + 1) + (1 + z)\frac{dq}{dz},
\end{equation}
where: \(j\) is the jerk parameter, \(a(t)\) represents the scale factor of the universe as a function of time, \(\dot{a}\) represents the first derivative of the scale factor, \(\ddot{a}\) represents the second derivative of the scale factor, \(\dddot{a}\) represents the third derivative of the scale factor, \(z\) represents redshift. In the context of the Taylor expansion of the scale factor around a reference time \(t_0\), the expansion takes the form:
\begin{equation}
\begin{aligned}
    \frac{a(t)}{a_0} &= 1 + H_0(t - t_0) - \frac{1}{2}q_0H_0^2(t - t_0)^2 + \frac{1}{6}j_0H_0^3(t - t_0)^3\\
    &+ O\left[(t - t_0)^4\right].
\end{aligned}
\end{equation}
where the subscript \(0\) denotes current values. Here, \(H_0\) represents the Hubble constant at the reference time \(t_0\), \(q_0\) is the deceleration parameter, and \(j_0\) is the jerk parameter at the same reference time. 
In contemporary cosmology, the jerk parameter has played a crucial role in refining our knowledge of the universe's evolution.\\\\\
\begin{figure}[!htb]
   \begin{minipage}{0.49\textwidth}
     \centering
   \includegraphics[scale=0.42]{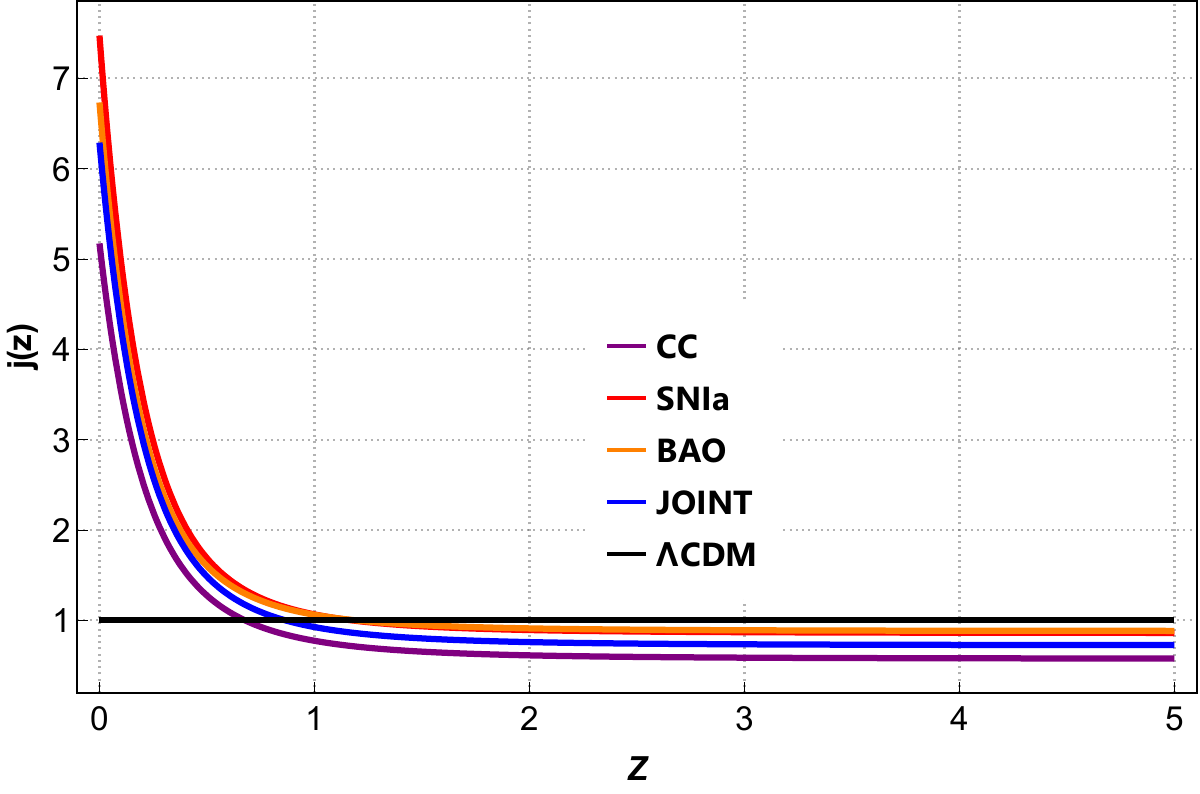}
\caption{Evolution of the jerk parameter with respect to redshift across various datasets, depicted alongside the $\Lambda$CDM model represented by the black line, characterized by $\Omega_{\mathrm{m0}}=0.3$ and $\Omega_\Lambda=0.7$.}\label{fig_10}
   \end{minipage}
\end{figure}

\section{Statefinder diagnostic}\label{sec9}
In the field of cosmology, the study of the universe's evolution, demands a comprehensive understanding of dark energy (DE) and its influence on cosmic expansion. To analyze this cosmic dynamic without bias toward any specific DE model, cosmologists employ a valuable tool known as the statefinder diagnostic parameter. This mathematical tool, pioneered by researchers \cite{state1,state2,state3,stste4}, employs higher derivatives of the cosmic scale factor to characterize the universe's expansion. Its primary purpose is to distinguish and compare different DE models effectively. What sets the statefinder diagnostic apart is its model-independent nature, enabling the exploration of various cosmological scenarios, including those with diverse forms of dark energy. The statefinder diagnostic is encapsulated in a parameter pair, denoted as $\{r,s\}$. These parameters are defined as follows:
\begin{equation}
r=\frac{\dddot{a}}{aH^{3}}, \quad s=\frac{r-1}{3\left( q-\frac{1}{2}\right)}.
\end{equation}
These parameters leverage higher-order derivatives of the scale factor, the Hubble parameter \(H\), and the deceleration parameter \(q\) to provide insights into cosmic expansion. Various possibilities in the $\{r,s\}$ and $\{q,r\}$ planes are exhibited to depict the temporal evolution of various DE models. With the assistance of the statefinder diagnostics pair. In these cases, some specific pairs typically correlate to classic DE models
such as $\{r,s\}=\{1,0\}$ represents $\Lambda $CDM model and $\{r,s\}=\{1,1\}$ indicates standard cold dark matter Model (SCDM) in FLRW background. Also,  $(-\infty ,\infty )$ yields static Einstein Universe. In the $r-s$ plane, $%
s>0$ and $s<0$ define a quintessence-like model and phantom-like model of the DE, respectively. Moreover, the evolution from phantom to quintessence can be observed by deviation from ${r,s}={1,0}$. On the other hand, $\{q,r\}=\{-1,1\}$ corresponds to the $\Lambda$CDM model while $\{q,r\}=\{0.5,1\}$ shows SCDM model. It is important to note that on a $r-s$
plane if the DE model's trajectories deviate from these standard values, the resulting model differs from the normal cosmic model.

\begin{figure}[!htb]
   \begin{minipage}{0.49\textwidth}
     \centering
   \includegraphics[scale=0.42]{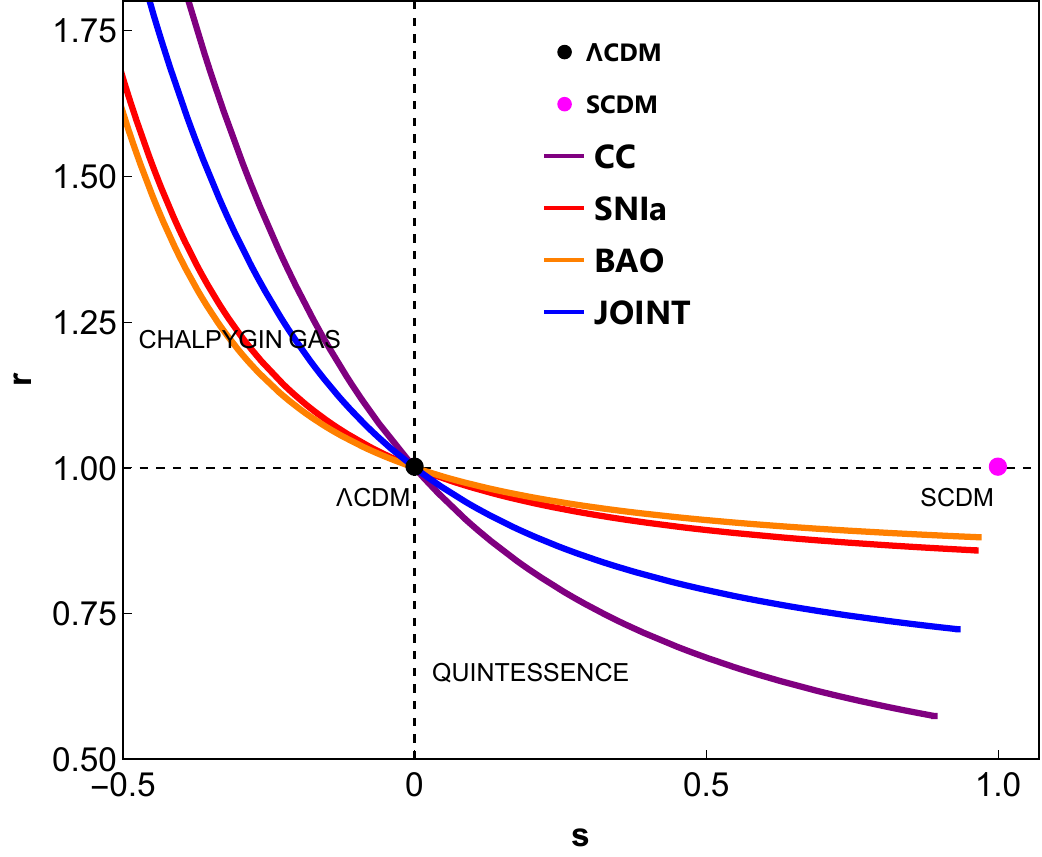}
\caption{Evolution of the $\{r, s\}$ profile with respect to redshift across various datasets.}\label{fig_12}
   \end{minipage}
\end{figure}
\begin{figure}[!htb]
   \begin{minipage}{0.49\textwidth}
     \centering
    \includegraphics[scale=0.42]{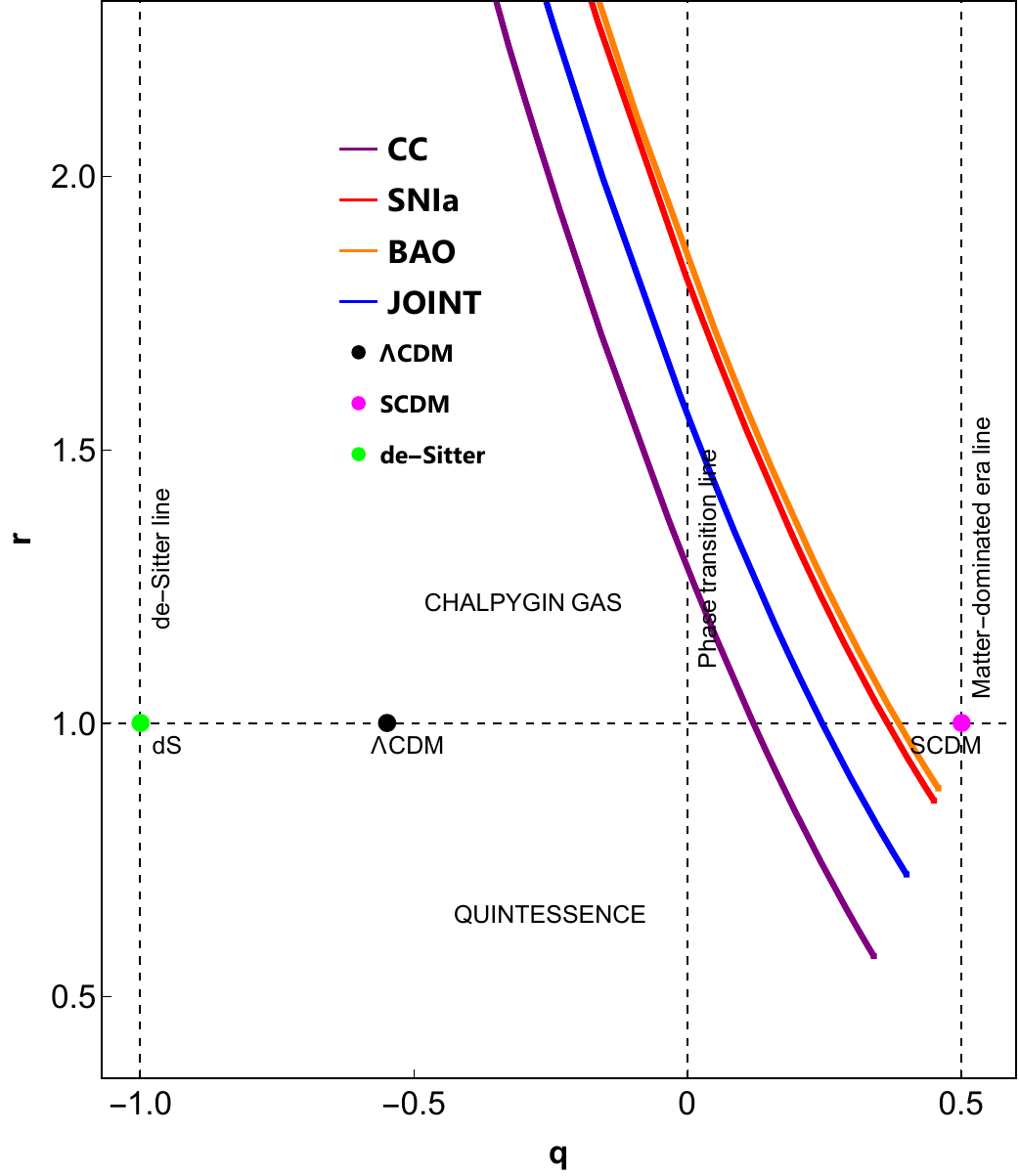}
\caption{Evolution of the $\{r, q\}$ profile with respect to redshift across various datasets.}\label{fig_13}
   \end{minipage}
\end{figure}

\section{Om Diagnostic}\label{sec10}
In the realm of cosmology, the $Om(z)$ diagnostic \cite{Om1,Om2,Om3,Om4} parameter plays a pivotal role in unraveling the cosmic mysteries. It's an essential tool for gauging the relative contribution of matter to the universe's total energy density, shedding light on the universe's overall geometry. $\omega_{m}$, the heart of this diagnostic, tells us about the ratio of current matter density to the critical density required for a flat universe. If $\omega_{m}$ is less than 1, it hints at an open universe, while a value greater than 1 suggests a closed one. This parameter serves as a geometric yardstick and plays a crucial role in testing the $\Lambda$CDM model, which posits three primary components: dark matter, ordinary matter (baryonic), and enigmatic dark energy. One of the remarkable aspects of $\omega_{m}$ is its ability to discern different dark energy models from the standard $\Lambda$CDM model. By altering the slope of $Om(z)$, this diagnostic efficiently distinguishes between various models. A positive slope signifies a quintessence model, while a negative slope points to a phantom model. A constant slope, on the other hand, aligns with the cosmological constant. This feature allows us to explore the intricate balance among these cosmic components. Much like the statefinder diagnostic, $Om(z)$ serves as a powerful tool for understanding cosmic evolution, geometry, and the interplay of matter in shaping the universe's dynamics. In a flat universe, $Om(z)$ is defined as:
\begin{equation}
Om(z)=\frac{\left( \frac{H(z)}{H_{0}}\right) ^{2}-1}{(1+z)^{3}-1}\text{.}
\label{34}
\end{equation}
\begin{figure}[!htb]
   \begin{minipage}{0.49\textwidth}
     \centering
   \includegraphics[scale=0.4]{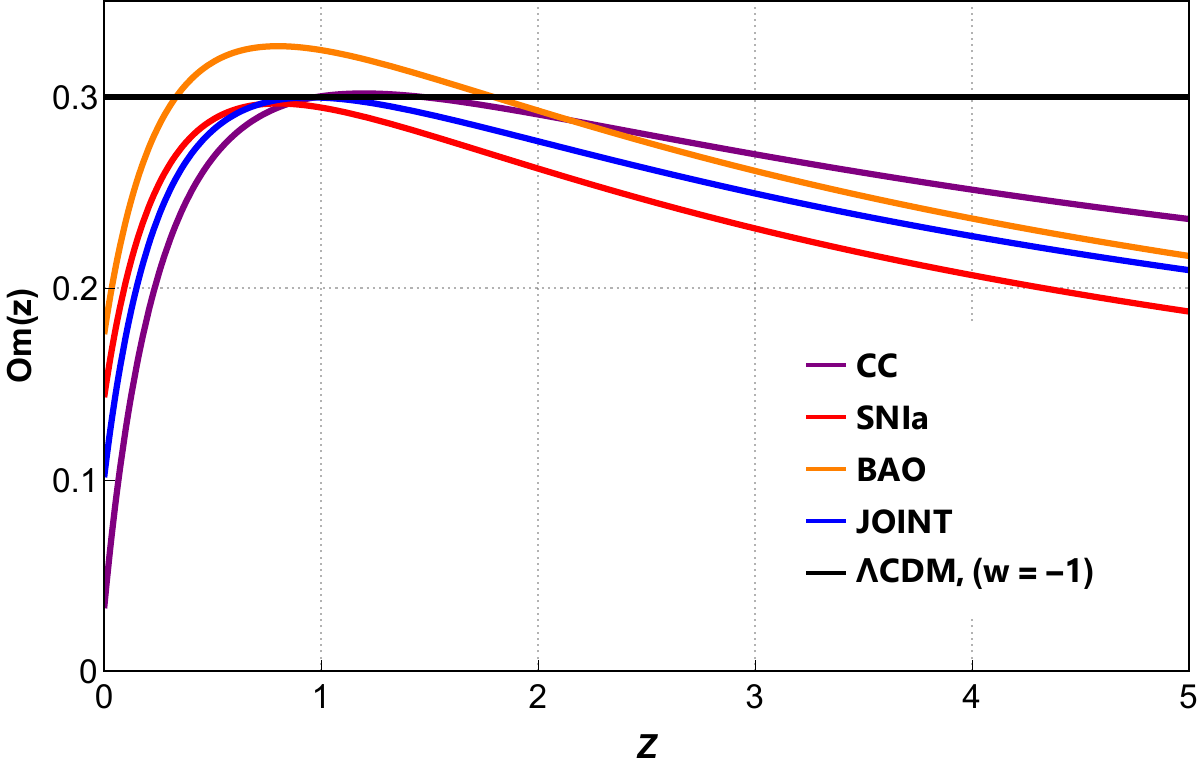}
\caption{Evolution of the $Om(z)$ profile with respect to redshift across various datasets.}\label{fig_144}
   \end{minipage}
\end{figure}
\section{Physical Parameters}\label{sec11}
In cosmology, physical parameters such as pressure, equations of state, and the density parameter (denoted as $\rho$) play crucial roles in describing the behavior and evolution of the universe. Here's a general overview of these concepts along with their associated formulas:
\subsection{Pressure Density $p$}
Pressure density influences the overall behavior of the universe's expansion. The pressure associated with cosmic components, such as matter, radiation, and dark energy, affects how they contribute to the expansion rate. Positive pressure (like that of matter) tends to slow down the expansion, while negative pressure (like that of dark energy) can accelerate it. The following figure FIG. 14 shows the transition from early positive values to the late negative values showing the accelerating expansion of the late universe. It is well described the early positive pressure contributes to the structure formation in the universe.
\begin{figure}[!htb]
   \begin{minipage}{0.49\textwidth}
     \centering
   \includegraphics[scale=0.4]{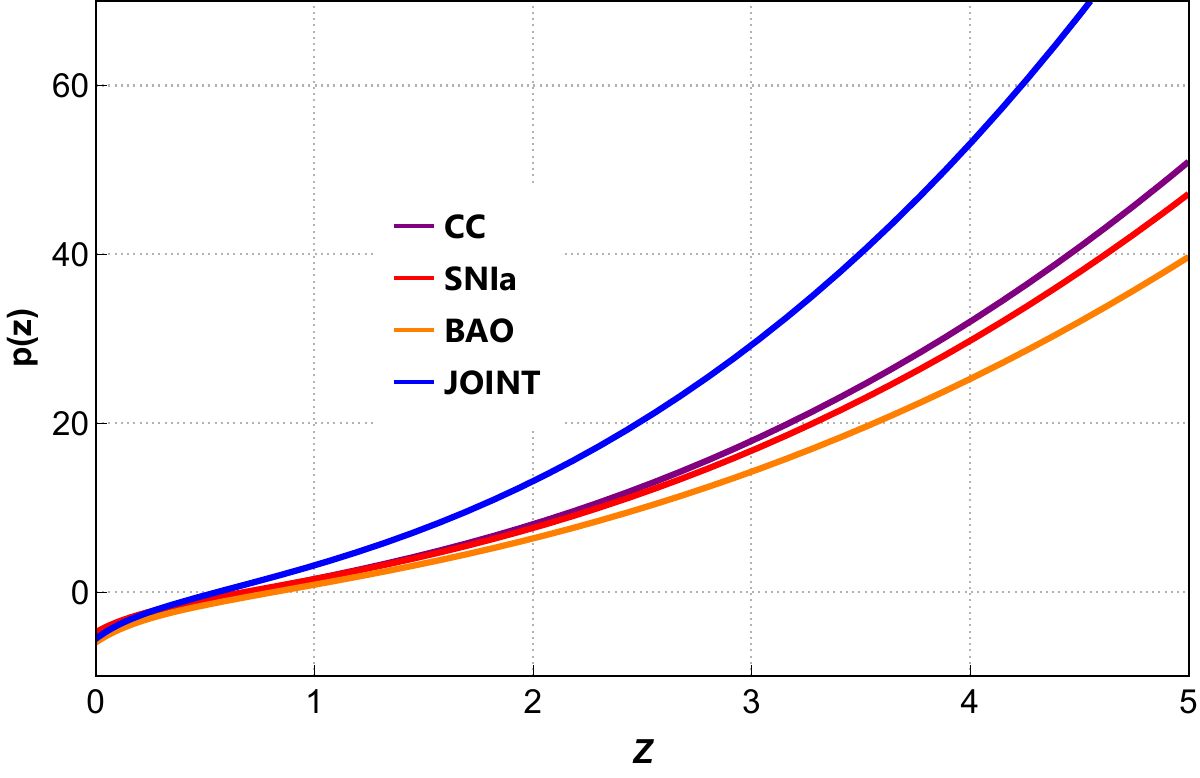}
\caption{Evolution of the Pressure density with respect to redshift across various datasets. }\label{fig_14}
   \end{minipage}
\end{figure}
\subsection{Energy Density $\rho$}
Energy density refers to the amount of energy contained within a given volume of space. In cosmology, it quantifies the total energy content of a specific cosmic component. Energy density is a crucial parameter as it determines the gravitational effects of that component on the overall cosmic dynamics. From the following figure FIG. 15, we can interpret the earlier high density universe to the present low values as expected.
\begin{figure}[!htb]
   \begin{minipage}{0.49\textwidth}
     \centering
   \includegraphics[scale=0.4]{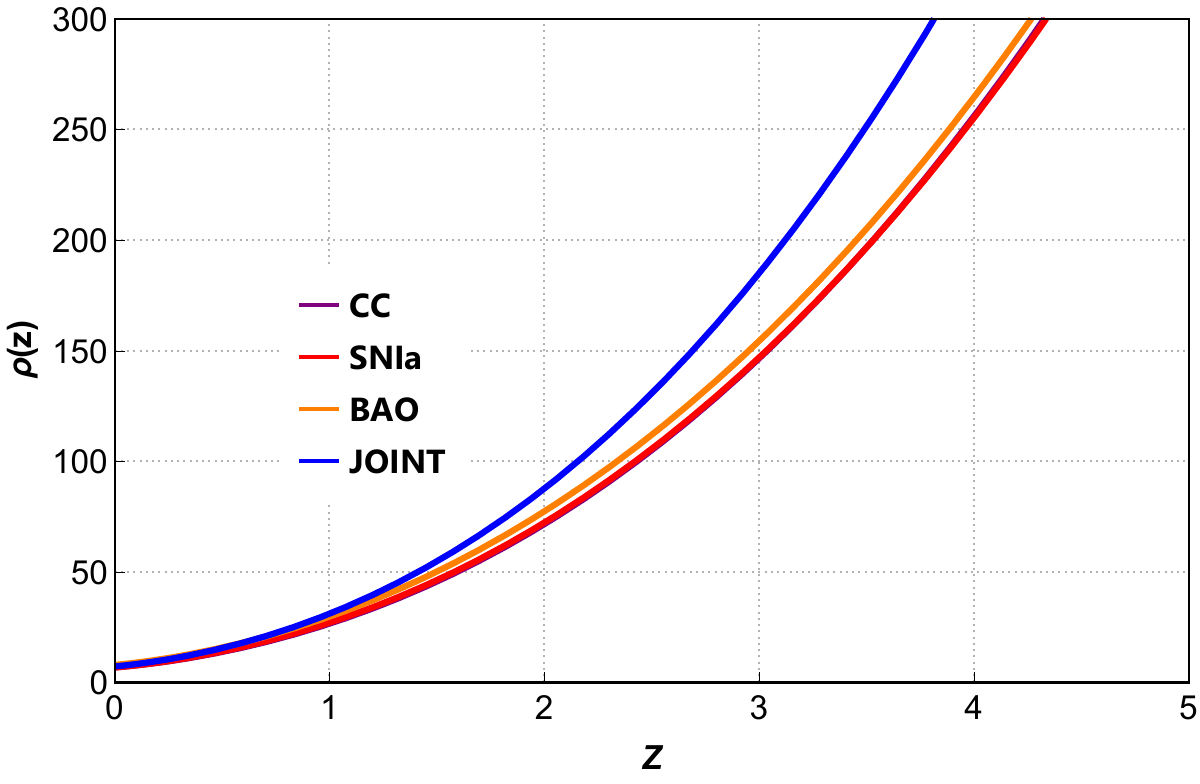}
\caption{Evolution of the Energy density with respect to redshift across various datasets.}\label{fig_15}
   \end{minipage}
\end{figure}
\subsection{Equation of State $\omega$}
The equation of state in cosmology establishes a relationship between the pressure (P) and energy density ($\rho$) of a cosmic component, effectively describing how pressure and density interrelate. In its general form, this equation is expressed as \(P = \omega \rho\), where "($\omega$)" serves as the parameter characterizing the equation of state. Different values of "($\omega$)" correspond to distinct cosmic components. For instance, when \(w = 0\), it corresponds to non-relativistic matter (e.g., cold dark matter). A value of \(\omega = 1/3\) corresponds to radiation (e.g., photons), while \(\omega \approx -1\) aligns with dark energy, exhibiting properties akin to a cosmological constant ($\Lambda$). These varying values of "($\omega$)" play a critical role in characterizing the behavior and properties of different cosmic components within the framework of cosmological models. The following figure FIG. 16 shows the evolution of the equation of state parameter in the due course of time. The universe enters into the quintessence phase in the recent past and remain in this phase in the near future too. The early positive values of $\omega$ shows the matter dominated era in the universe.

\begin{figure}[!htb]
   \begin{minipage}{0.49\textwidth}
     \centering
   \includegraphics[scale=0.4]{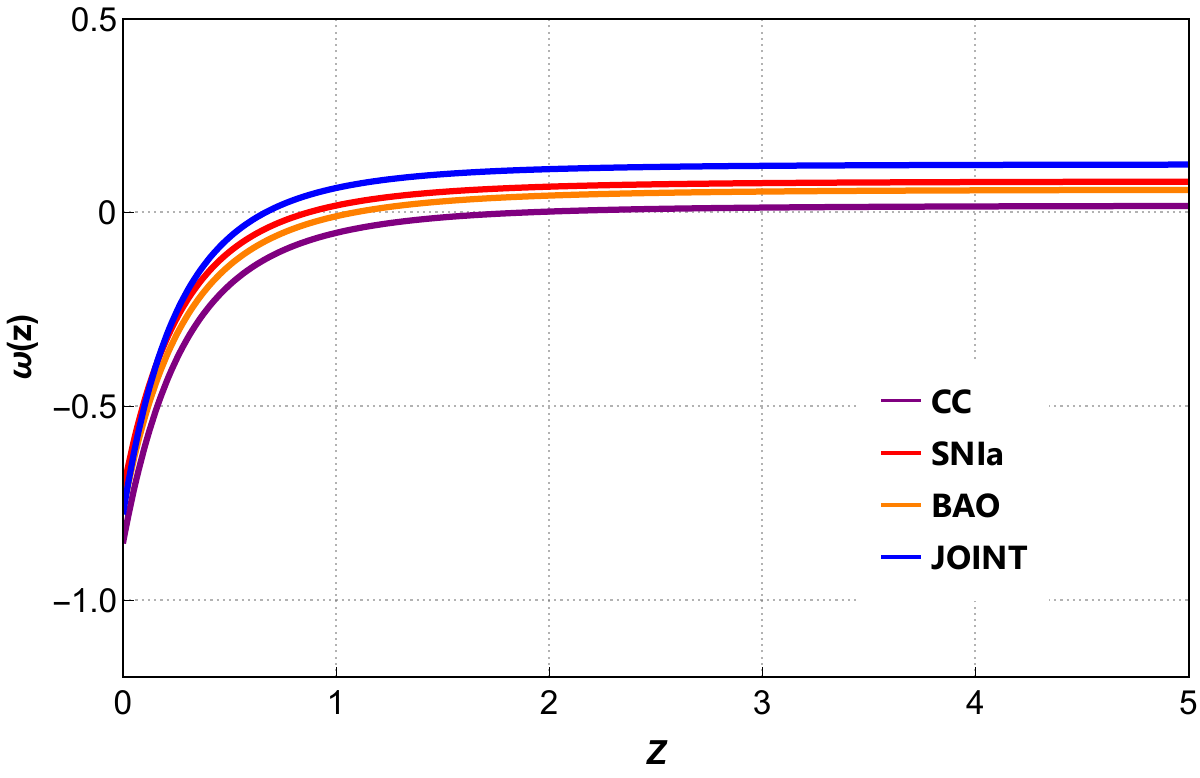}
\caption{Evolution of the EoS Parameter with respect to redshift across various datasets.}\label{fig_16}
   \end{minipage}
\end{figure}
\section{Statistical Analysis}\label{sec12}
To determine the optimal model for our analysis, it is crucial to consider the number of free parameters associated with each model, in addition to the $\chi_{\text{min}}^{2}$ value obtained. While choosing among the various information criteria available in the literature is not a straightforward task \cite{liddle2004}, we opt for the most commonly used ones. One of these is the Akaike Information Criterion (AIC) \cite{schwarz1978,liddle2004,nesseris2013}, defined as:
\begin{equation}\label{eq62}
AIC \equiv -2 \ln \mathcal{L}_{\text{max}} + 2p_{\text{tot}} = \chi_{\text{min}}^{2} + 2p_{\text{tot}}    
\end{equation}
Here, $p_{\text{tot}}$ represents the total number of free parameters in the specific model, and $\mathcal{L}_{\text{max}}$ signifies the maximum likelihood of the model being considered. Additionally, we incorporate the Bayesian Information Criterion (BIC), introduced by \cite{schwarz1978,liddle2004,nesseris2013}, which is defined as:
\begin{equation}\label{eq63}
BIC \equiv -2 \ln \mathcal{L}_{\text{max}} + p_{\text{tot}} \ln \left(N_{\text{tot}}\right)    
\end{equation}
By utilizing the definitions \eqref{eq62} and \eqref{eq63}, we compute the discrepancies $\triangle A I C$ and $\triangle B I C$ relative to the $\Lambda$CDM model in question. In accordance with \cite{jeffreys1998theory}, if $0 < |\triangle A I C| \leq 2$, it implies that the compared models may be viewed as compatible with each other. Conversely, if $|\triangle A I C| \geq 4$, it indicates that the model with the higher AIC value is not supported by the data. Similarly, for $0 < |\triangle B I C| \leq 2$, the model exhibiting the higher BIC value is marginally less favored by the data. In cases where $2 < |\Delta B I C| \leq 6$ ($|\triangle B I C|>6$), the model with the higher BIC values is significantly (highly) less favored. The specific distinctions among the investigated cosmological models are detailed in \ref{tab_AIC}. \\\\
In cosmology, the terms "P-value" and "L-statistic" are commonly used in statistical analyses to assess the significance of observations and test hypotheses.
\begin{itemize}
    \item P-value (Probability Value)
    The P-value is a statistical measure that quantifies the evidence against a null hypothesis. It tells you the probability of observing data as extreme or more extreme than what you have, assuming that the null hypothesis is true. In cosmology, P-values are often used in the context of hypothesis testing \cite{p1,p2,p3}. For example, when analyzing cosmic microwave background (CMB) data, cosmologists may calculate P-values to assess whether the observed CMB power spectrum matches the predictions of a particular cosmological model. A low P-value suggests that the data is inconsistent with the model, while a high P-value suggests that the data is consistent with the model.

    \item L-statistic (Likelihood Statistic)
    The L-statistic, often referred to as the likelihood ratio \cite{L1,L2,L3}, is a statistical measure used to compare the likelihood of observing data under two different hypotheses. In cosmology, the L-statistic is frequently used when comparing different cosmological models or parameter values. For instance, in Bayesian cosmological analyses, the likelihood function quantifies how well a given model explains the observed data. The L-statistic is used to compare the likelihoods of different models, and models with higher likelihoods are considered more compatible with the data.
\end{itemize}
In practical terms, when cosmologists perform statistical analyses in cosmology, they often calculate likelihoods and P-values to evaluate the goodness of fit between observational data and theoretical models. These statistical tools help cosmologists make inferences about the parameters of the universe, such as the density of dark matter, dark energy properties, and the geometry of the universe. 
\begin{figure}[!htb]
   \begin{minipage}{0.49\textwidth}
     \centering
   \includegraphics[scale=0.46]{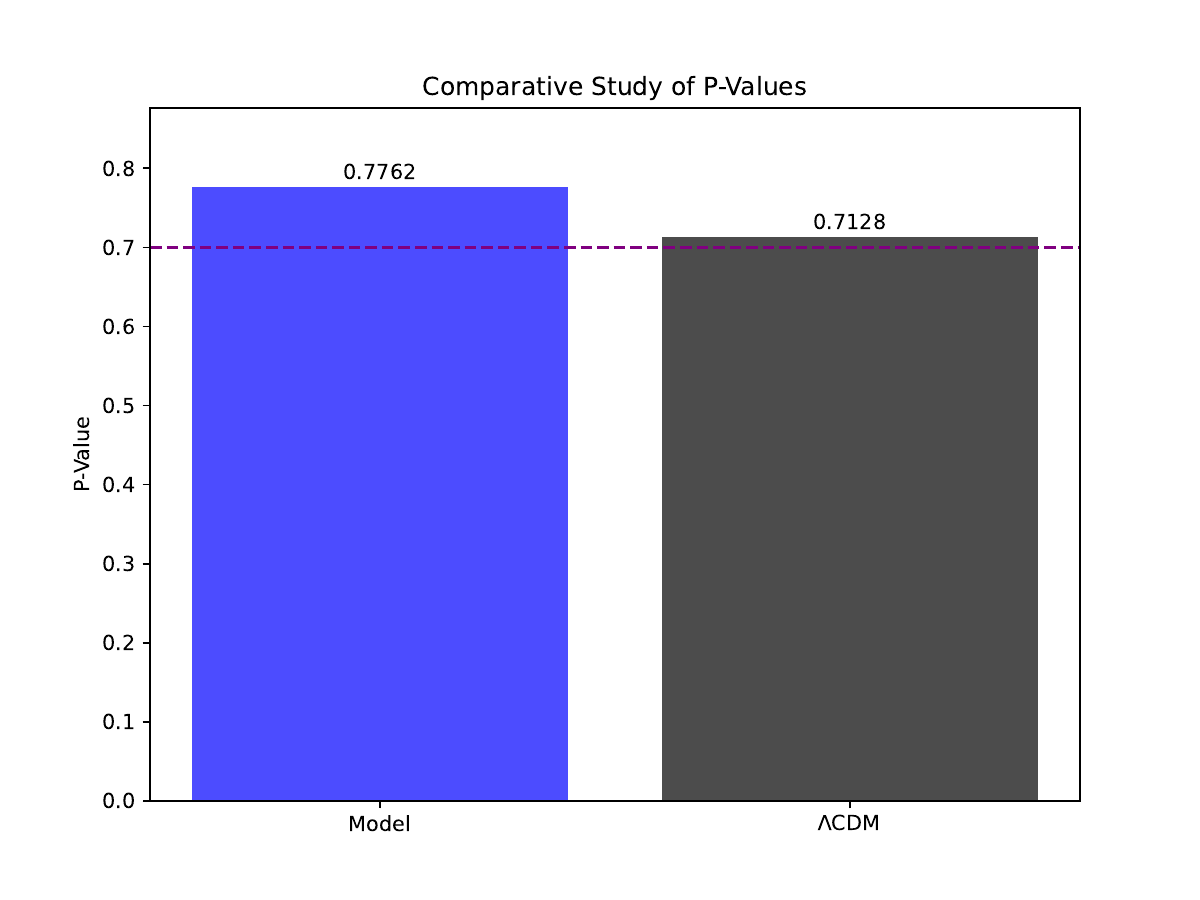}
\caption{The comparative Comparative Study of P-Values of joint dataset}\label{fig_18}
   \end{minipage}
\end{figure}
\begin{figure}[!htb]
   \begin{minipage}{0.49\textwidth}
     \centering
   \includegraphics[scale=0.46]{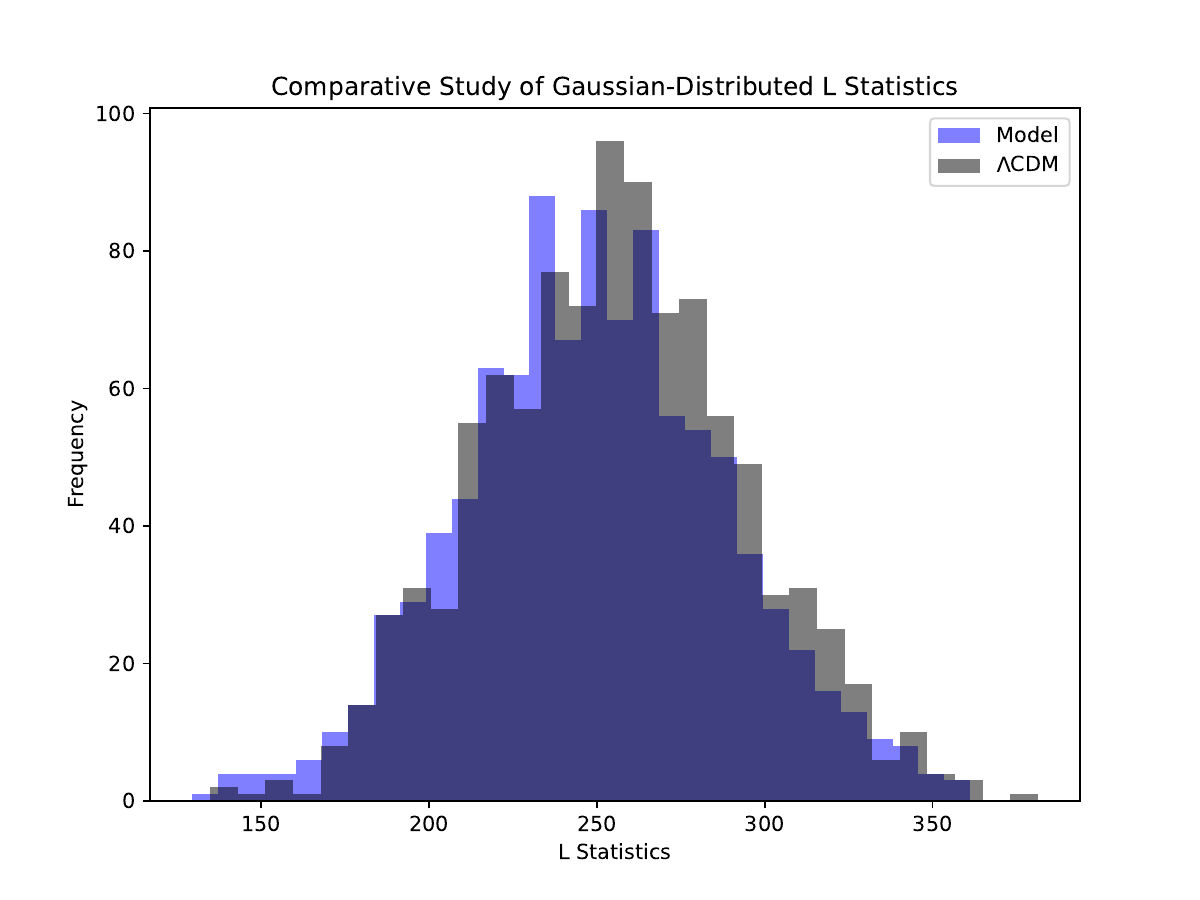}
\caption{The comparative histogram plot of the distribution of L statistics for both models of joint dataset}\label{fig_19}
   \end{minipage}
\end{figure}
\begin{figure}[!htb]
   \begin{minipage}{0.49\textwidth}
     \centering
   \includegraphics[scale=0.46]{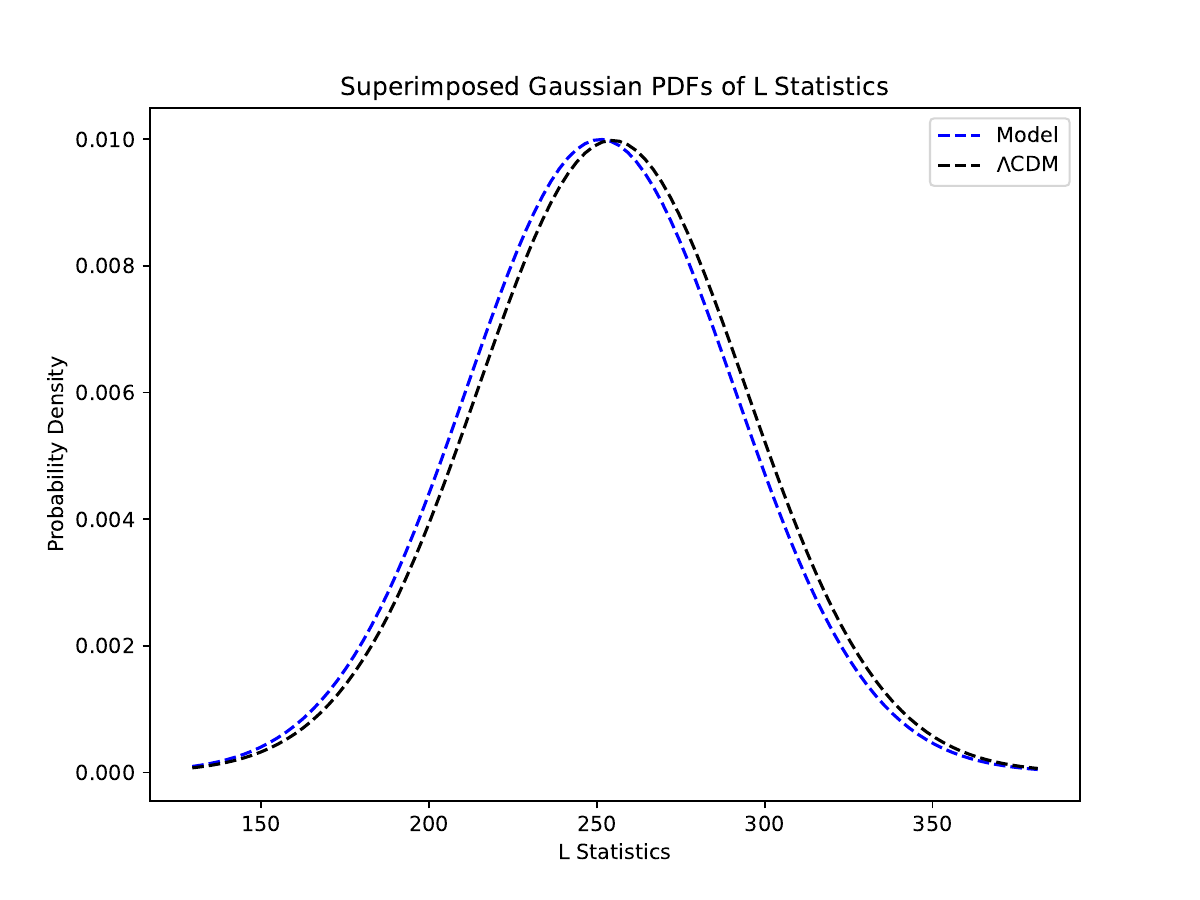}
\caption{The superimposed overlays of Gaussian Probability Density Functions (PDFs) on the L statistics for both models of joint dataset.}\label{fig_20}
   \end{minipage}
\end{figure}
\begin{table*}
\begin{center}
{\begin{tabular}{|c|c|c|c|c|c|c|c|c|c|c|c|}
\hline
{\bf Model} & Dataset & ${\chi_{\text{tot}}^2}^{min} $ & $\chi_{\text {red }}^2$ &$\mathcal{K}_{\textrm{f}}$ & $A I C_c$ & $\Delta A I C_c$ &{\bf$B I C$} & {\bf $\Delta B I C$} & P-value & L-statistic  \\[0.1cm]
\hline
$\Lambda$CDM Model & CC & 15.5126& 0.9053 & 3& 21.5126 & 0 & 25.8146 & 0 & 0.560256 & 26.140535\\[0.1cm]
& SNIa & 1036.50 & 0.9356 & 3 & 1042.51 & 0 & 1057.43 & 0 & 0.989391 & 203.623840\\[0.1cm]
& BAO & 26.4431& 0.9182&3& 32.4431 & 0& 34.9427&0&0.550075 &10.811795 \\[0.1cm]
& Joint & 1107.35  & 0.9645 & 3& 1113.35 & 0 & 1128.83 & 0 & 0.712755 & 254.538726 \\[0.1cm]
\hline
Model &CC& 15.6871 & 0.8953 &3 & 21.6871& 0.17& 25.9891 & 0.17& 0.554023& 26.140535\\[0.1cm] 
& SNIa &1033.91& 0.9298 & 3 & 1039.91 & -2.59 & 1054.84 & -2.29 & 0.991874 & 203.635102\\[0.1cm]
& BAO & 27.3445& 0.9177 & 3 & 33.3445  & 0.90 & 35.8441 & 0.90 & 0.664331 & 9.446289 \\[0.1cm]
& Joint &1108.32 & 0.9634  & 3 & 1114.32  & 0.97 & 1129.80 & 0.97 & 0.776201 & 251.087532 \\[0.1cm] \hline
\end{tabular}}
\caption{Summary of ${\chi_{\text{tot}}^2}^{min} $, $\chi_{\text {red }}^2$, $A I C_c$, $\Delta A I C_c$, {\bf $B I C$, $\Delta B I C$}, P-value and L-statistic  for $\Lambda$CDM and $f(Q,T)$  model across all dataset.}\label{tab_AIC}
\end{center}
\end{table*}

\newpage
\section{Results}\label{sec13}
\paragraph{deceleration parameter} Fig \ref{fig_9} represents the redshift evolution of the deceleration parameter ($q$) across various cosmological epochs, denoted by the redshift ($z$), as constrained by different datasets, including Cosmic Chronometers (CC), Type Ia Supernovae (SNIa), Baryon Acoustic Oscillations (BAO), and a Joint Dataset encompassing CC + SNIa + Gamma-Ray Bursts (GRB) + Quasars (Q) + BAO and $\Lambda$CD paradigm. At high redshifts, corresponding to earlier epochs in the universe's history, the model exhibits a noteworthy convergence toward a consistent value of $q$ across all analyzed datasets, specifically around $0.2763$. As the universe evolves and redshift decreases, it undergoes a phase transition from a decelerating phase to an accelerating one. Nonetheless, the long-term behavior of the two models starkly diverges. An intriguing and noteworthy observation is that the model predicts an acceleration at half-time higher than the $\Lambda$CDM model with an approximate value of $q(0)\approx -1$, whereas the $\Lambda$CDM model predicts a value of approximately $q(0)\approx -0.5$. This deceleration parameter maintains consistency across all the datasets employed in our comprehensive analysis.\\\\
\paragraph{jerk parameter}
Fig \ref{fig_10} illustrates the redshift-dependent behavior of the jerk parameter across the various datasets. At high redshifts, the value of $j$ in the proposed model is across all datasets and the $\Lambda$CDM model consistently converges to approximately 1. However, important differences do exist at lower redshifts, with a very significant difference appearing at $z=0$, where the proposed model predicts a value higher than the $\Lambda$CDM value of  $j =1$ across all datasets. Determining the current value of $j$ through observations can be a crucial test for the proposed model cosmological model.\\\\
\paragraph{$\{r, s\}$ profile}
The $\{r, s\}$ profile of our obtained cosmological model, as depicted in Fig \ref{fig_12}, reveals a fascinating cosmological evolution. At the initial stages, the model exhibits values in the range $r < 1$ and $s > 0$, which correspond to characteristics akin to the quintessence region In cosmology, quintessence is a dynamic form of dark energy with a varying energy density and negative pressure, potentially explaining the universe's accelerated expansion. Further by crossing intermediate $\Lambda$CDM fixed point $\{0,1\}$, it transitions into a different region where $r > 1$ and $s < 0$, represent Chaplygin gas. Chaplygin gas is a cosmological model featuring exotic matter with unique properties, used to explain cosmic expansion. Remarkably, this $\{r, s\}$ profile consistently exhibits the same behavior with respect to all the datasets used in our analysis.\\\\
\paragraph{$\{r, q\}$ profile}
The $\{r, q\}$ profile of our derived cosmological model, as shown in Fig \ref{fig_13}, reveals a fascinating evolution. Initially, the model exhibits values within the region $q > 0$ and $r < 1$, which resembles characteristics commonly linked with the quintessence. As the Universe evolves, a distinctive transition occurs, leading the model into a different region characterized by $q < 0$ and $r > 1$, represented by the Chaplygin gas scenario. Notably, this $\{r, q\}$ profile consistently demonstrates the same behavior across all the datasets integrated into our analysis.\\\\
\paragraph{Om Diagnostic}
Figure \ref{fig_144} illustrates the evolution of $Om(z)$ at various redshifts (z) for our model across different datasets. Notably, for the CC, SNIa, and Joint datasets, $Om(z)$ consistently remains below the current matter density parameter $\Omega_{m0}$. This observation suggests that the model consistently resides in the phantom region throughout the entire evolution of the Universe at all redshifts. On the other hand, for the BAO dataset, $Om(z)$ consistently remains below the current matter density parameter for redshifts $z > 1.8$. This suggests that the model falls within the phantom domain. As the Universe evolves, $Om(z)$ transitions to being higher than the current matter density parameter in the redshift range of $0.34 > z > 1.8$, indicating that the model enters the quintessence domain. Finally, $Om(z)$ decreases below the current matter density parameter for redshifts $z < 0.34$, signifying that the model reverts to the phantom domain.\\\\
\paragraph{Physical Parameters}
Figures \ref{fig_14}, \ref{fig_15}, and \ref{fig_16} provide the necessary illustration into the dynamic evolution of physical parameters, pressure ($p$), energy density ($\rho$), and the equation of state parameter ($\omega$) at different redshifts. These figures meticulously portray how these parameters change as a function of redshift, with each curve representing a distinct set of model parameters derived from diverse datasets. The beauty of these plots lies in their ability to unveil the intricate phases of evolution exhibited by these physical parameters at different points in cosmic history. In particular, Figure \ref{fig_14} unveils the transition in the pressure parameter: it shifts from being positive in the early universe to becoming increasingly negative as time progresses, hinting at the onset of late-time cosmic acceleration. However, what truly sets this observation apart from conventional cosmological models like the $\Lambda$CDM model is the intriguing revelation that the pressure doesn't just remain negative. Meanwhile, Figure \ref{fig_15} charts the path of energy density ($\rho$), illustrating how it initially diminishes from its higher values in the past to its present, smaller magnitude. Figure \ref{fig_16} provides a glimpse into the cosmic temperament of the Universe through the equation of state parameter. For the region \( z > 1 \), the model's \( \omega \) falls within \( 0 < w < \frac{1}{3} \), aligning with the behavior expected in the Matter-dominated era, where pressure is negligible compared to energy density. As the Universe evolves and enters the epoch \( z < 1 \), the \( \omega \) value shifts into the range \( -1 < w < -\frac{1}{3} \). This shift signifies a transition to a quintessence field domain, suggesting the presence of a dynamic form of dark energy with negative pressure, contributing to the observed cosmic acceleration. We can witness the matter dominated era in the early universe to the late dark energy-dominated era in the quintessence realm.\\\\
\paragraph{Statistical Analysis}
Based on the table \ref{tab_AIC} furnishes a detailed comparison between the $f(Q,T)$ model and the standard $\Lambda$CDM model based on various statistical indicators. Starting with ${\chi_{\text{tot}}^2}^{min}$, which represents the minimum total chi-square, the $f(Q,T)$ model yields slightly higher values across all datasets compared to the $\Lambda$CDM model. This suggests that, when attempting to minimize the discrepancies between the model and observational data, $\Lambda$CDM achieves a marginally better fit. The reduced chi-square ($\chi_{\text{red}}^2$) accounts for the number of degrees of freedom, providing a more normalized measure. Here, too, the $f(Q,T)$ model exhibits slightly higher values, indicating that $\Lambda$CDM offers a slightly improved fit when considering the complexity of the models. Moving on to information criteria, $\Delta A I C_c$ and $\Delta B I C$ are crucial for model comparison. In both metrics, positive values for the $f(Q,T)$ model signify a weaker degree of support compared to $\Lambda$CDM. Typically, $\Delta AIC$ values in the range of $(0,2)$ suggest strong support, while values in $(4,7)$ are less supported. Values beyond $10$ indicate models without any meaningful support. Therefore, the positive $\Delta A I C_c$ values for the $f(Q,T)$ model suggest that, in terms of information criteria, $\Lambda$CDM is more strongly supported. Considering P-values and L-statistic, which provide insights into the overall goodness of fit, the $f(Q,T)$ model generally yields higher P-values and lower L-statistic values compared to $\Lambda$CDM. This implies that the $f(Q,T)$ model has a comparatively poorer fit to the observational data. Fig. \ref{fig_19} provides the comparative histogram distribution of L statistics for $\Lambda$CDM model and our Model. The $x$-axis represents the L statistics values, while the $y$-axis represents the frequency or count of data points within each bin. You can observe how the L statistics are distributed for both models. Look for differences in the central tendencies (peaks) and spreads (widths) of the distributions. Since our model's histogram is similar to the $\Lambda$CDM histogram, it suggests that our model is a reasonable fit for the data, Fig. \ref{fig_20} provides the superimposed plot overlays Gaussian Probability Density Functions (PDFs) on the L statistics for both models. The $x$-axis represents the L statistics values, and the $y$-axis represents the probability density. Again the Gaussian PDF of our model is closely aligned with the histogram of L statistics, it suggests a better match between the $\Lambda$CDM model and the data.
\section{Conclusion}\label{sec14}
In this study, we conducted a comprehensive and rigorous examination of the $f(Q, T)$ cosmological model. We extend the analysis conducted in classical gravity by incorporating a parametrization of the deceleration parameter, as outlined in \cite{sofuouglu2023observational}. Our investigation encompasses the realm of $f(Q,T)$ gravity, allowing us to explore its implications and effects on the parametrization. This model was compared against a range of cosmological observations, encompassing 31 Cosmic Chronometers, 1071 type Ia supernovae measurements, 162 Gamma Ray Bursts (GRBs), 24 measurements from compact radio quasars, 17 Baryon Acoustic Oscillation (BAO) measurements, and CMB distant prior. To determine the best-fit value for the model parameters, we employed the Markov Chain Monte Carlo (MCMC) methodology, allowing us to derive the optimal fit for these parameters. Subsequently, utilizing these best-fit values, the data-fitting process yielded exceptionally accurate outcomes for both the CC and SNIa datasets. We observed the redshift evolution of the deceleration parameter ($q$) across different cosmological epochs using various datasets, the model exhibits a consistent convergence toward a value of approximately $0.2763$ at high redshifts, indicative of the early universe's behavior. As the universe evolves and redshift decreases, a transition from a decelerating to an accelerating phase occurs. Notably, the proposed model predicts an acceleration at a rate higher than the $\Lambda$CDM model, with an approximate value of $q(0) \approx -1$, while the $\Lambda$CDM model predicts a value of approximately $q(0) \approx -0.5$. Recent studies have provided varied estimates for the present-day deceleration parameter (\(q(0)\), deviating from the standard \(\Lambda\)CDM model. In the cited research \cite{ryan2021constraints}, the authors found that the value of $q(0)$ is approximately \(0.05641\). Similarly, in the study \cite{Modelindependent}, a different estimate of \(q(0) \approx 0.84331\) was derived. In the referenced research \cite{akarsu2012cosmological}, \(q(0) \approx -0.73\). In the study \cite{kumar2012observational}, the authors obtained \(q(0) \approx -0.34\). In the investigation by Xu et al. \cite{xu2008constraints}, three models were introduced, each suggesting distinct values. In Model 1, \(q(0) \approx -0.657\), in Model 2, \(q(0) \approx -0.982\), and in Model 3, \(q(0) \approx -0.410\). These non-standard values indicate a departure from the expected $\Lambda$CDM prediction and emphasize the growing recognition of variations in the cosmological dynamics, prompting further investigation into alternative models and theoretical frameworks. This divergence is maintained consistently across all datasets utilized in our comprehensive analysis. Moving to the jerk parameter, at high redshifts, both the proposed model and the $\Lambda$CDM model converge to a value of approximately 1 across all datasets. However, significant differences emerge at lower redshifts, particularly at $z=0$, where the proposed model consistently predicts a higher value for $j$ compared to the $\Lambda$CDM value of $j=1$ across all datasets. The determination of the current value of $j$ through observations emerges as a crucial test for validating the proposed cosmological model. This presents a crucial test for the proposed model. The analysis of the $\{r, s\}$ profile revealed a fascinating cosmological evolution. Our model exhibited characteristics resembling quintessence and Chaplygin gas phases, maintaining consistency across all datasets. The $\{r, q\}$ profile depicted a dynamic evolution, transitioning from quintessence-like behavior to a Chaplygin gas scenario. This behavior remained consistent across all datasets. The $Om(z)$ analysis indicated that our model consistently positions $Om(z)$ below \(\Omega_{m0}\) for CC, SNIa, and Joint datasets, indicating a sustained presence in the phantom region. The observed transition in the BAO dataset from the phantom to quintessence domains introduces complexity to the model's dynamic evolution. The analysis of pressure ($p$), energy density ($\rho$), and the equation of state parameter ($\omega$) across varying redshifts reveals deviations from standard cosmological models like the $\Lambda$CDM model. The EoS reveals a nuanced cosmic temperament, shifting from matter domination to quintessence, offering a unique perspective distinct from conventional models. The comparative analysis between our proposed model and the $\Lambda$CDM Model across various statistical metrics, including chi-square values, information criteria, and goodness-of-fit indicators, the results consistently favor the $\Lambda$CDM model over the $f(Q,T)$ model. The statistical analyses collectively suggest that $\Lambda$CDM provides a slightly better balance between goodness of fit and model. In conclusion, while our proposed model provides intriguing insights and predicts unique cosmic behaviors, such as super acceleration, the standard $\Lambda$CDM Model remains favored by statistical criteria. The differences observed in various cosmological parameters and profiles underscore the complexity of cosmological models and the need for ongoing research to refine our understanding of the universe's evolution. Further observations and refined datasets may shed more light on these intriguing cosmological phenomena.\\\\
\textbf{Acknowledgement: }Author SKJP thanks IUCAA, Pune for hospitality and other facilities under its IUCAA associateship program, where a large part of work has been done. G. Mustafa is very thankful to Prof. Gao Xianlong from the Department of Physics, Zhejiang Normal University, for his kind support and help during this research. Further, G. Mustafa acknowledges Grant No. ZC304022919 to support his Postdoctoral Fellowship at Zhejiang Normal University, China. \\\\






\bibliographystyle{unsrt}
\bibliography{mybib}

\end{document}